\documentclass[seceq,preprint]{ptptex}

\usepackage{epsfig, color,graphicx}
\usepackage{amsmath}
\usepackage{amssymb}
\newcommand{\nn}{\nonumber \\}
\newcommand{\bea}{\begin{eqnarray}}
\newcommand{\ena}{\end{eqnarray}}

\newcommand{\hs}[1]{\hspace{#1 mm}}
\renewcommand{\a}{\alpha}

\renewcommand{\c}{\gamma}
\renewcommand{\d}{\delta}

\newcommand{\s}{\sigma}

\newcommand{\la}{\lambda}

\newcommand{\p}[1]{(\ref{#1})}
\newcommand{\tm}{\tilde{m}}
\newcommand{\tr}{\tilde{r}}

\newcommand{\squaret}{\kern1pt\vbox{\hrule height 0.9pt\hbox{\vrule width
0.9pt\hskip 2pt\vbox{\vskip 5.5pt}\hskip 3pt\vrule width 0.3pt}\hrule height
0.3pt}\kern1pt}
\begin{document}

\preprintnumber[3cm]{
KU-TP 033}
\title{\large  Black Holes in the Dilatonic Einstein-Gauss-Bonnet Theory
in Various Dimensions IV \\
-- Topological Black Holes with and without Cosmological Term --
}
\author{
Nobuyoshi {\sc Ohta}$^{a,}$\footnote{e-mail address: ohtan at phys.kindai.ac.jp}
and Takashi {\sc Torii}$^{b,}$\footnote{e-mail address: torii at ge.oit.ac.jp}
}

\inst{
$^a$Department of Physics, Kinki University, Higashi-Osaka,
Osaka 577-8502, Japan\\
$^b$Department of General Education, Osaka Institute of Technology,
Asahi-ku, Osaka 535-8585, Japan
}
\abst{
We study black hole solutions in the Einstein gravity with Gauss-Bonnet term,
the dilaton and a positive ``cosmological constant'' in various dimensions.
Physically meaningful black holes with a positive cosmological term
are obtained only for those  in static spacetime with $(D-2)$-dimensional hyperbolic space
of negative curvature and $D\geq 5$.
We construct such black hole solutions of various masses numerically in $D=5,6$
and 10 dimensional spacetime and discuss their properties.
In spite of the positive cosmological constant the spacetime approach
anti-de Sitter spacetime asymptotically. The black hole solutions  exist
for a certain range of the horizon radius, i.e., there are lower and 
upper bounds for the size of black holes.  
We also argue that it is quite plausible that there is no black hole solution
for hyperbolic space in the case of no cosmological constant.
}

\maketitle

\section{Introduction}

This is the fourth and final of a series of our papers in which
we study the static black hole solutions in dilatonic Einstein-Gauss-Bonnet theory
in higher dimensions.~\cite{GOT1,GOT2,OT3}

There have been many works on black hole solutions in dilatonic gravity
since the work in Refs.~\citen{GM} and~\citen{GHS}.
Because there are higher-order quantum corrections and dilaton in string
theories,~\cite{GS} it is important to study how these modify the results.
Several works have studied the effects of higher order terms but only in four
dimensions,~\cite{KMRTW,AP,TYM,CGO1,CGO2} or considered theories without
dilaton.~\cite{BD,GG,Cai}
Motivated by this situation, we started our study of black hole solutions and
their properties in the theory with the higher order corrections and dilaton
in dimensions higher  than or equal to four.
The simplest higher order correction is the Gauss-Bonnet (GB) term,
which may appear in heterotic string theories.

In our first paper in this series,~\cite{GOT1} we have studied black hole solutions
with the GB correction term and dilaton without the cosmological term for asymptotically
flat solutions in various dimensions from 4 to 10 with $(D-2)$-dimensional
hypersurface with curvature signature $k=+1$.
We tried to find topological black holes for $k=-1$, but it turned out
that there is no solution without the cosmological term.
We have then presented our results on black hole solutions with a negative
cosmological term with $(D-2)$-dimensional hypersurface with
$k=0, \pm 1$ in the second and third papers.~\cite{GOT2,OT3}
In the context of string theories, it is very interesting to examine asymptotically
non-flat black hole solutions with possible applications to AdS/CFT and dS/CFT
correspondence in mind.~\cite{CNOS}
Discussions of the origin of such cosmological terms are given in Refs.~\citen{AGMV,POL}.
In our systematic study of these   static black hole solutions, those cases with the
positive cosmological term and those for hyperbolic space without the cosmological term
have not been studied. In this paper, we continue our study of the black holes
for these remaining cases and complete our study.
Cosmological solutions in such a system are considered in Ref.~\citen{BGO}.

This paper is organized as follows. In \S~2, we first summarize our action
with the GB and the cosmological terms, and give basic equations to solve.
We then discuss symmetry properties of the theory which will be
useful in our following analysis.
In \S~3, boundary conditions at the horizon and asymptotic behaviors are
discussed for our solutions.
In \S~4, we discuss asymptotic expansions and allowed parameter regions for
the existence of black holes in the case with the positive cosmological term.
We show that there is no allowed parameter region for $D=4$ and hence
there is no black hole solution.
We then argue that there is no asymptotically de Sitter (dS) solutions due to
the presence of a non-normalizable mode in higher dimensions. We find, however, that
there are asymptotically anti-de Sitter (AdS) solutions for appropriate range of
parameters even though we have the positive cosmological term.
In \S~5, we present these solutions in $D=5,6$ and 10 with a positive cosmological
term for $k=-1$. Though we do not present explicit solutions for
other $D=7, 8$ and 9, the properties are similar~\cite{GOT1} and we expect
that there are similar solutions.
In \S~6, we show that there is no black hole solution for $k=0$.
In \S~7, we argue that there is no physically meaningful black hole solution
for $k=1$ for some parameter range in our action in general unless
miraculous cancellation of the growing mode of the dilaton happens,
although there does exist an exact regular dS solution with constant dilaton.
For other range of the parameters, there may be regular (not black hole)
solutions, which will be studied elsewhere.
These solutions and discussions are given for particular choices of the parameters
in our theory, but we expect that the qualitative behaviors do not change
for other choices.

In \S~8, we discuss topological black holes with $k=-1$ for zero cosmological term,
and show that there is no black hole solution in $D=4$. In higher dimensions,
we find that there is a narrow range of the value of the dilaton at the horizon
where solutions may exist. We have not found, however,  black hole solutions
even for this range, and most probably there is no black hole solution at least
for the dilaton coupling $\c=\frac12$, though we do not have rigorous proof.

\S~9 is devoted to conclusions and discussions.
In particular, we summarize our main results obtained in this series of papers.

\section{Dilatonic Einstein-Gauss-Bonnet theory}

\subsection{Action and basic equations}

We consider the following low-energy effective action for a
heterotic string
\bea 
S=\frac{1}{2\kappa_D^2}\int d^Dx \sqrt{-g} \left[R - \frac12
 (\partial_\mu \phi)^2
 + \a_2 e^{-\c \phi} R^2_{\rm GB} -\Lambda e^{\la \phi} \right],
\label{act}
\ena 
where $\kappa_D^2$ is a $D$-dimensional gravitational constant,
$\phi$ is a dilaton field, $\alpha_2=\a'/8$ is a numerical
coefficient given in terms of the Regge slope parameter $\a'$,
and
$R^2_{\rm GB} = R_{\mu\nu\rho\sigma} R^{\mu\nu\rho\sigma}
- 4 R_{\mu\nu} R^{\mu\nu} + R^2$ is the GB correction.
For the moment, we leave the coupling constant of dilaton $\gamma$ arbitrary,
while the ten-dimensional critical string theory predicts $\c=1/2$
which is the value we choose in our numerical analysis.
We have also included the positive cosmological constant $\Lambda=(D-1)_2/\ell^2$
with possible dilaton coupling $\la$.
If this is the only potential of the dilaton field, there is no stationary point and the dilaton
cannot have a stable asymptotic value. However, for asymptotically
(A)dS solutions, the GB term can be regarded as an additional ``potential"
in the asymptotic region, and we will see that it is possible to have
the solutions where the dilaton takes finite constant value at infinity.
There may be several possible sources of ``cosmological terms'' with different
dilaton couplings, so we leave $\lambda$ arbitrary and specify it
in the numerical analysis.

We parametrize the metric as
\bea
ds_D^2 = - B e^{-2\d} dt^2 + B^{-1} dr^2 + r^2 h_{ij}dx^i dx^j,
\ena
where $h_{ij}dx^i dx^j$ represents the line element of a
$(D-2)$-dimensional hypersurface with constant curvature
$(D-2)(D-3)k$ and volume $\Sigma_k$ for $k=\pm 1,0$.

The metric function $B=B(r)$ and the lapse function $\d=\d(r)$ depend only on the
radial coordinate $r$. The field equations are~\cite{GOT2,OT3}
\bea
&& \bigl[(k-B)\tr^{D-3}\bigr]' \frac{D-2}{\tr^{D-4}}h -\frac12 B \tr^2 {\phi'}^2
 - (D-1)_4\,e^{-\c\phi}\frac{(k-B)^2}{\tr^2} \nn
&& \hs{10} + 4(D-2)_3\, \c e^{-\c\phi}B(k-B)(\phi''-\c {\phi'}^2) \nn
&& \hs{10} + 2(D-2)_3\,\c e^{-\c\phi}\phi'\frac{(k-B)[(D-3)k-(D-1)B]}{\tr}
-\tr^2 \tilde \Lambda e^{\la \phi}= 0\,,
\label{fe1} \\
&& \delta'(D-2)\tr h + \frac12 \tr^2 {\phi'}^2
 -2(D-2)_3\, \c e^{-\c\phi}(k-B)(\phi''-\c {\phi'}^2) =0 \,,
 \label{fe2} \\
&&
(e^{-\d} \tr^{D-2} B \phi')' = \c (D-2)_3 e^{-\c\phi-\d} \tr^{D-4}
\Big[ (D-4)_5 \frac{(k-B)^2}{\tr^2} + 2(B'-2\d' B)B' \nn
&& \hs{10} -4(k-B)BU(r)
-4\frac{D-4}{\tr}(B'-\d'B)(k-B) \Big] + e^{-\d} \tr^{D-2} \la \tilde\Lambda e^{\la\phi},
\label{fe3}
\ena
where we have defined the dimensionless variables: $\tr \equiv r/\sqrt{\a_2}$,
$\tilde \Lambda \equiv \a_2 \Lambda$, and the primes in the field equations
denote the derivatives with respect to $\tr$. Namely we measure our length
in the unit of $\sqrt{\a_2}$. We have defined
\bea
(D-m)_n &\equiv& (D-m)(D-m-1)(D-m-2)\cdots(D-n), \nn
\label{h-def}
h &\equiv& 1+2(D-3) e^{-\c\phi} \Big[ (D-4) \frac{k-B}{\tr^2}
 + \c \phi'\frac{3B-k}{\tr}\Big], \\
\label{tilh-def}
\tilde h &\equiv& 1+2(D-3) e^{-\c\phi} \Big[(D-4)\frac{k-B}{\tr^2}
+\c\phi'\frac{2B}{\tr} \Big], \\
U(r) &\equiv& (2 \tilde h)^{-1} \Bigg[ (D-3)_4 \frac{k-B}{\tr^2 B}
-2\frac{D-3}{\tr}\Big(\frac{B'}{B}-\d'\Big) -\frac12 \phi'^2 \nn
&& + (D-3)e^{-\c\phi} \Bigg\{ (D-4)_6 \frac{(k-B)^2}{\tr^4 B}
- 4 (D-4)_5 \frac{k-B}{\tr^3}\Big(\frac{B'}{B}-\d'-\c\phi'\Big) \nn
&& -4(D-4)\c \frac{k-B}{\tr^2}\Big( \c \phi'^2 +\frac{D-2}{\tr}\phi'-\Phi \Big)
+8 \frac{\c\phi'}{\tr} \biggl[\Big(\frac{B'}{2}-\d' B\Big)\Big(\c\phi'-\d'
+\frac{2}{\tr} \Big) \nn
&& -\frac{D-4}{2\tr}B' \biggr] + 4(D-4)\Big(\frac{B'}{2B}-\d' \Big)
\frac{B'}{\tr^2}-\frac{4\c}{\tr}\Phi (B'-2\d'B)\Bigg\}
-\frac{1}{B} \tilde \Lambda e^{\la \phi}\Biggr],\\
\Phi &\equiv& \phi'' +\Big(\frac{B'}{B}-\d' +\frac{D-2}{\tr}\Big) \phi'.
\label{dil}
\ena

\subsection{Symmetry and scaling}

It is useful to consider several scaling symmetries of our field equations
(or our model).
Firstly the field equations are invariant under the transformation:
\bea
\gamma \to -\gamma, ~~
\lambda \to -\lambda, ~~
\phi \to -\phi\, .
\label{sym1}
\ena
By this symmetry, we can restrict the parameter range of $\c$ to $\c\geq 0$.

The field equations~\p{fe1} -- \p{fe3} have a shift symmetry:
\bea
\phi \to \phi-\phi_{\ast}, ~~
\tilde\Lambda \to e^{(\la-\c)\phi_{\ast}} \tilde\Lambda, ~~
r \to e^{-\gamma\phi_{\ast}/2}r.
\label{sym2}
\ena
where $\phi_{\ast}$ is an arbitrary constant.
This changes the magnitude of the cosmological constant. Hence this may
be used to generate solutions for different cosmological constants,
given a solution for some cosmological constant and $\tr_H$.
Details will be discussed in \S~5.

The third one is the shift symmetry under
\bea
\delta \to \delta - \delta_{\ast}, ~~
t \to  e^{-\delta_{\ast}}t,
\label{sym3}
\ena
with an arbitrary constant $\delta_{\ast}$, which may be used to shift
the asymptotic value of $\delta$ to zero.

For $k=0$, the field equations~\p{fe1}--\p{fe3} are
invariant under the scaling transformation
\bea
B \to a^2 B, ~~
\tr \to a \tr,
\ena
with an arbitrary constant $a$.
If a black hole solution with the horizon radius $\tr_H$ is obtained,
we can generate solutions with different horizon radii but the same $\tilde \Lambda$
by this scaling transformation.

The model \p{act} has several parameters of the theory such as
$D$, $\alpha_2$, $\Lambda$, $\gamma$, and $\lambda$. The black hole solutions
have also physical parameters such as the horizon radius $\tr_H$ and
the value of $\delta$ at infinity.
However owing to the above symmetries (including the scaling by $\alpha_2$),
we can reduce the number of the parameters and are left only with
$D$, $\gamma\geq 0$, $\lambda$, and $\tr_H$.

\section{Boundary conditions}
\label{BC}

In this paper, we study solutions with spherical ($k=+1$), planar ($k=0$) and
hyperbolic ($k=-1$) symmetries with a positive cosmological constant $\tilde \Lambda >0$
except for \S~8 where we study solutions with $k=-1$ and $\Lambda=0$.
In this section, we discuss the boundary conditions and asymptotic behaviors
for these solutions.

\subsection{Regular horizon}

Let us first examine the boundary conditions of the black hole spacetime.
We impose the following boundary conditions for the metric functions:
\begin{enumerate}
\item[1.]
The existence of a regular black hole horizon at $\tr=\tr_{BH}$:
\bea
\label{hor}
B(\tr_{BH})=0, ~~
B'(\tr_{BH})>0, ~~
|\d(\tr_{BH})| < \infty, ~~
|\phi(\tr_{BH})|< \infty\, .
\ena
\item[2.]
The absence of singularities outside the black hole event horizon ($\tr > \tr_{BH}$).
\end{enumerate}

Since our model (\ref{act}) has the positive cosmological constant,
there may be cosmological horizon at $\tr=\tr_{CH}>\tr_{BH}$ with
the second condition in Eq.~(\ref{hor}) replaced by $B'(\tr_{CH})<0$.
Here and in what follows, the values of various quantities at the black hole and
the cosmological horizons are denoted with subscripts $BH$ and $CH$, respectively.
When we refer to both horizons, we simply use the subscript $H$.

When  $B'(\tr_{BH})=0$, the horizon is degenerate, and the spacetime can be the
extreme black hole if $B''(\tr_{BH})>0$.

At the horizon, it follows from Eqs.~\p{fe1}, \p{h-def}, and \p{tilh-def} that
\bea
&& B_H=0, \nn
&&
h_H = 1+(D-4 -\c\phi_H' \tr_H)C,\nn
&&
\tilde h_H= 1+(D-4)C,\nn
&&
B_{H}' h_H = \frac{(D-3)k}{\tr_H} +\frac{(D-4)_5kC}{2} - \frac{\tr_H L}{D-2},
\label{bhor}
\ena
where we have defined
\bea
C\equiv \frac{2(D-3)ke^{-\c\phi_H}}{\tr_H^2},~~
L \equiv e^{\la\phi_H} \tilde\Lambda.
\label{c}
\ena
Using these in Eq.~\p{fe3} gives
\bea
&&
2C\c\Big[-2  L \tr_H^2  \Big\{ 1 + C \big[D-4 -(D-2) \c \la\big]
+  (D-2)C^2 \c \big[(D-6) \c -(D-4) \la \big] \Big\} \nn
&& \hs{10}
+  (D-2)k \Big\{ 2(D-3) + (D-4)(3 D-11) C \nn
&& \hs{15}
+ (D-4) C^2 \big[ (D-2)(3D-11) \c^2+ (D-4)_5 \big] +2(D-2)_5 C^3 \c^2 \Big\}\Big]
\tr_H^2 \phi_H'^2 \nn
&&
+ 2\Big[4C^2 \c ^2L^2 \tr_H^4
+ 2 L \tr_H^2 \Big\{ 1+ 2C\big[D-4-(D-2)\c\la\big]
\nn
&& \hs{20}
-C^2\big[ 2(D-2)(2D-5)\c^2
+4(D-2)(D-4)\c\la-(D-4)^2 \big]
\nn
&& \hs{20}
-2(D-2) (D-4) C^3 \c\big[(D-2)\c+(D-4)\la \big] \Big\} \nn
&&  \hs{10}
+(D-2) k\Big\{  (D-1)_2 (D-4) C^2 \c^2\big[ 2+2kC-(D-4)_5 C^2 \big] \nn
&& \hs{30}
- \big[1+(D-4)C \big]^2 \big[2(D-3)+(D-4)_5 C \big]\Big\} \Big]\tr_H\phi_H' \nn
&&
+ 4 \frac{C}{k}\c L^2 \tr_H^4 + 4 (D-2) L \tr_H^2 \Big\{ \la \big[ 1+(D-4)C \big]^3 \nn
&& \hs{30}
+ C\c \big[ D(D-4)^2 C^2 + (D-4)(D+1) C -(D-2) \big] \Big\} \nn
&& \hs{5}
+ (D-1)_2 (D-2) kC\c \Big\{ 2(D-2) -4 (D-4)C - (D-4)^2 (D+1) C^2 \Big\}=0,
\label{pder}
\ena
This equation gives the value of the derivative of the dilaton field at a horizon
for given $\tr_H$ and $\phi_H$.

\subsection{Asymptotic behavior at infinity and cosmological horizon}

If the solution can be continued outside the cosmological horizon, we impose
the condition that the leading term of the metric function
$B$ comes from dS radius at infinity.
Since the curvatures do not vanish for asymptotically non-flat case in general
and the GB term contributes in the asymptotic region, the spacetime can
approach non-dS spacetime as we will see later.
So we impose the following condition:
\begin{enumerate}
\item[3.]
``Appropriate" asymptotic behavior at infinity ($\tr \to \infty$):
\bea
\label{as}
B \sim  \tilde{b}_2 \tr^2 +k - \frac{2\tilde M}{\tr^{\mu}}, ~~~
\d(r) \sim \d_0 + \frac{\d_1}{\tr^{\s}}, ~~
\phi \sim \phi_0 + \frac{\phi_1}{\tr^{\nu}} \,,
\ena
with finite constants $\tilde{b}_2$, $\tilde M$, $\d_0$, $\d_1$,  $\phi_0$, $\phi_1$
and positive constant $\mu$, $\s$, $\nu$.
Depending on the sign of the constant $\tilde{b}_2$, the spacetime has different
asymptotics, i.e. dS for $\tilde{b}_2<0$, flat for $\tilde{b}_2=0$,
and AdS for  $\tilde{b}_2>0$.
\end{enumerate}

\subsection{Behaviors of dilaton between black hole and cosmological horizons}

The equation of the dilaton field (\ref{fe3}) is rewritten as
\begin{eqnarray}
e^{-\delta}B\phi^{\prime \prime }
+\frac{1}{\tr^{D-2}}\bigl(\tr^{D-2}e^{-\delta}B \bigr)' \phi'
= e^{-\delta}\bigl(\gamma e^{-\gamma\phi} \tilde R_{GB}^2
+\lambda \tilde{\Lambda}e^{\lambda\phi} \bigr).
\label{horizoneq}
\end{eqnarray}
where
\bea
&&\tilde{R}_{GB}^2=\frac{(D-2)_3}{\tr^2}
\biggl[
(D-4)_5 \frac{(k-B)^2}{\tr^2} + 2(B'-2\d' B)B'
\nn
&& \hspace{10mm}
-4(k-B)\Bigl(\frac12 B''-B\delta''-\frac32 B'\delta'+B\delta'^2\Bigr)
-\frac{4(D-4)}{\tr}(k-B)(B'-\d'B) \biggr].
\ena
In the Einstein theory ($\tilde{R}_{GB}^2=0$),
since $B=0$ on the  horizons, $\phi'_H$ is expressed by $\phi_H$ as
\begin{equation}
\phi_H'= \frac1{B_H'}\lambda \tilde{\Lambda}e^{\lambda\phi_H}.
\label{derp}
\end{equation}
We assume $\lambda<0$ from the consideration of the asymptotic condition~\p{as}
presented in the next section.
This implies that $\phi$ decreases around the black hole because
$B'$ is positive near the black hole horizon, i.e., the right hand side (r.h.s.)
of Eq.~\p{derp} is negative. Similarly it is locally increasing function
around the cosmological horizon because $B_{CH}'<0$.
These mean that the dilaton field has to have a local minimum
in $\tr_{BH}<\tr<\tr_{CH}$.
On the other hand, at the extremum of the dilaton field ($\phi'=0$),
Eq.~\p{horizoneq} gives
\begin{eqnarray}
B\phi^{\prime \prime }
=\lambda \tilde{\Lambda}e^{\lambda\phi}.
\end{eqnarray}
It then follows that $\phi^{\prime \prime}<0$
since $B>0$ for $\tr_{BH}<\tr<\tr_{CH}$. Hence the dilaton field does
not have the local minimum in this region. This contradicts the
behaviors around both the horizons. We can thus conclude that there
is no static
black hole solution with $k=\pm 1,\: 0$ in the Einstein-dilaton case.

In the GB case, however, the situation is different.
The r.h.s. of Eq.~\p{horizoneq} can take both signs
at the horizons and at the extremum of the dilaton field.
Hence Eqs.~\p{fe1}-\p{fe3} may have the consistent solutions between
$\tr_{BH}$ and $\tr_{CH}$ which satisfy the condition at each horizon.
It should be noted here that $\phi$ and $\phi'$ cannot be free but are related
with each other by the condition~\p{pder} at the cosmological horizon. In order to
satisfy this condition, we have to tune the value of $\phi_{BH}$ as a shooting
parameter for each $\tr_{BH}$. This will have important consequence in
the subsequent discussions.

\section{Asymptotic expansions and allowed parameter regions}

\subsection{Asymptotic expansion}

Substituting Eqs.~(\ref{as}) into the field equations (\ref{fe1}) and
(\ref{fe3}), one finds the conditions that the leading terms ($\tr^2$ and constant
terms in each equation) should satisfy are given by
\bea
&& \label{phi_inf_1}
(D)_3\gamma\; e^{-\gamma\phi_0} \tilde{b}_2^{~2}
+ \lambda \tilde{\Lambda}e^{\lambda\phi_0} = 0 \,, \\
&& \label{B_inf_1}
(D-1)_4\;e^{-\gamma\phi_0} \tilde{b}_2^{~2}-(D-1)_2 \tilde{b}_2
-\tilde{\Lambda}e^{\lambda\phi_0} = 0 \,,
\ena
which determine $\tilde b_2$ and $\phi_0$,
while $\d_0$ can be arbitrary because only its derivative appears in our field
equations.
Since $\tilde{\Lambda}$ and $\gamma$ are positive, $\lambda$ should
be negative by Eq.~\p{phi_inf_1}. The positivity of $\tilde{\Lambda}$ means
that $\tilde b_2\ne 0$ by Eq.~\p{B_inf_1}. So there is no solution with
asymptotically flat spacetime under our assumptions.
We also find that
\bea
\left\{
\begin{array}{lcl}
\gamma> \dfrac{D-4}{D}|\lambda| & ~ \Longrightarrow ~ & \tilde{b}_2<0 ~~ ({\rm dS}),
\\
\gamma= \dfrac{D-4}{D}|\lambda| & ~ \Longrightarrow ~ & \tilde{b}_2=0 ~~
(\mbox{not allowed}),
\rule[0mm]{0mm}{8mm}
\\
\gamma< \dfrac{D-4}{D}|\lambda| & ~ \Longrightarrow ~ & \tilde{b}_2>0 ~~ ({\rm AdS}).
\rule[0mm]{0mm}{8mm}
\end{array}
\right.
\label{class}
\ena
Eqs.~(\ref{phi_inf_1}) and (\ref{B_inf_1}) give
\begin{eqnarray}
&&
\label{b2_inf}
\tilde{b}_2^{~2}
=\frac{-\lambda\tilde{\Lambda}}{(D)_3 \gamma}
\biggl[\frac{D(D-3)}{(D-1)_2}\frac{\tilde{\Lambda}\gamma}{(-\lambda)}
\biggl(1+\frac{(D-4)\lambda}{D\gamma}\biggr)^2
\biggr]^{\frac{\gamma+\lambda}{\gamma-\lambda}},
\\
&&
\label{phi_inf_2}
e^{\phi_0}
=\biggl[\frac{D(D-3)}{(D-1)_2}\frac{\tilde{\Lambda}\gamma}{(-\lambda)}
\biggl(1+\frac{(D-4)\lambda}{D\gamma}\biggr)^2
\biggr]^{\frac{1}{\gamma-\lambda}}.
\end{eqnarray}

The candidates of the next leading terms for Eqs.~\p{fe1}-\p{fe3} are
respectively given by
\begin{eqnarray}
&& \label{next1-2}
\bigl[\mu-(D-3)\bigr]\bigl[(D-4)\lambda -D\gamma\bigr]\tilde{M}\tr^{-\mu}
- 2 \gamma\lambda(\nu+1)\bigl[\nu-(D-1)\bigr]\tilde{b}_2\phi_1\tr^{2-\nu}
\,,\\
&& \label{next2-2}
\rule[0mm]{0mm}{7mm}
\bigl[(D-4)\lambda-D\gamma\bigr]\sigma \d_1 \tr^{-\sigma}
+ 2\lambda \gamma\nu(\nu+1)\phi_1 \tr^{-\nu} \,,\\
&&
\rule[0mm]{0mm}{7mm}
4(D-2)\lambda\gamma\sigma(\sigma-D)\tilde{b}_2\delta_1 \tr^{-\sigma}
\nonumber \\
&& \hs{10}
+\Bigl\{(D)_2\lambda\gamma(\lambda+\gamma)
+\bigl[(D-4)\lambda+D\gamma\bigr]\nu\bigl[\nu-(D-1)\bigr]\Bigr\}\tilde{b}_2\phi_1
\tr^{-\nu}
\nonumber \\
&& \hs{10}
+ 4(D-2)\lambda\gamma\bigl[\mu-(D-2)\bigr]\bigl[\mu-(D-3)\bigr]\tilde{M}\tr^{-\mu-2}
\,,
\label{next3-2}
\end{eqnarray}
which should all vanish.
There are two different classes which give consistent expansions.
One is  $\mu=D-3$ and $\nu, \s >D-1$, and we rename the coefficient $\tilde M$
as $\tilde M_0$.
The other class is $\mu=\nu-2=\sigma-2$ and
\bea
\nu =\nu_{\pm} \equiv \frac{D-1}2 \left[1 \pm \sqrt{1 - \frac{4(D)_2\la\c(\la-\c)
\bigl[(D-4)\la+D\c\bigr]}{(D-1)^2\bigl[(D-4)^2\la^2-D^2\c^2-8(D-1)_2\la^2\c^2\bigr]}}
\;\right].
\label{nu}
\ena

The asymptotic forms of the fields are then
\bea
&&
\phi \sim \phi_0
+ \frac{\phi_{+}}{\tr^{\nu_{+}}}
+ \cdots\,,\nn
&&
\label{behavior}
B \sim \tilde{b}_2 \tilde{r}^2 +k
- \frac{2 \tilde{M}_{+}}{\tr^{\nu_{+}-2}} - \frac{2\tilde{M}_0}{\tr^{D-3}} + \cdots\,,\\
&&
\delta \sim \frac{\delta_{+}}{\tr^{\nu_{+}}}+ \cdots \, .\nonumber
\ena
There will be $\phi_{-}/\tr^{\nu_{-}}$ term in the asymptotic
behavior of $\phi$  in general. Such mode is a non-normalizable mode.
Hence we
tune the boundary condition of $\phi_{BH}$ to kill this term.
Note that while $B$ has the term $r^{-\nu_{+}+2}$,
the $g_{tt}$ component of the metric behaves as
\begin{eqnarray}
-g_{tt} \sim \tilde b_2 \tr^2 + k - \frac{2\tilde{M}_0}{\tr^{D-3}}.
\end{eqnarray}
This value of $\tilde{M}_0$ is the gravitational mass of the black holes.
We will present our results in terms of this function.
Thus it is convenient to define the mass function $\tm_g(\tr)$ by
\bea
-g_{tt} = \tilde b_2 \tr^2 + k - \frac{2\tm_g(\tr)}{\tr^{D-3}}.
\label{mass}
\ena

\subsection{Allowed parameter regions}
\label{AllowedRegions}

We rewrite the indices $\nu_{\pm}$ as
\begin{eqnarray}
\nu_{\pm}= \frac{D-1}{2}\biggl[1\pm \sqrt{1-\frac{{\tilde{m}}^2}{{\tilde{m}}_{BF}^2}}
\biggr],
\label{nu-pm}
\end{eqnarray}
where the mass square of Breitenlohner and Freedman (BF) bound is defined by~\cite{BF}
\begin{eqnarray}
\tilde{m}_{BF}^2&&=-\frac{(D-1)^2}{4\tilde{\ell}_{\rm AdS}^2}
=-\frac{(D-1)^2}{4}\tilde{b}_2.
\end{eqnarray}
This expression is usually used in asymptotically AdS case.
We define the mass square of the dilaton field as
\begin{eqnarray}
\tilde{m}^2 = - \frac{(D)_2\la\c(\la-\c)\bigl[(D-4)\la+D\c\bigr]}
 {(D-4)^2\la^2-D^2\c^2-8(D-1)_2\la^2\c^2}\tilde{b}_2.
\label{mass1}
\end{eqnarray}
by the analogy with the discussion in BF bound.
This mass can be considered to be the value of
the second derivative of the effective potential of the dilaton field, and
the sign of the mass square defined by Eq.~\p{mass1} gives the information
about the shape of the effective potential. When $\tilde{m}^2>0$, the potential
has a local minimum around $\phi_0$ while it has a local maximum when
$\tilde{m}^2<0$.

For asymptotically dS case $\gamma> (D-4)|\lambda|/D$ (see Eq.~\p{class}),
we find that $\tilde{m}^2<0$
and $\nu_{-}<0$ which means that $\phi_{-}$ is a growing mode. This is consistent
with the picture that the dilaton field rolls down the potential slope for $B<0$.
If there is such a mode in the solutions, the dominant term in the metric function
$B$ is not $\sim r^2$ any more but  $\sim r^{2+|\nu_-|}$.
In order to satisfy the asymptotic condition \p{as}, we should eliminate
the growing mode $\phi_{-}$ by tuning $\phi_{BH}$.
However, the degree of freedom of $\phi_{BH}$ is already used to obtain the regular
cosmological horizon. So such a degree does not remain. There may be
a rare occasion where the conditions of the existence of regular cosmological
horizon and asymptotically dS behavior happen to be satisfied simultaneously for
a certain size of black hole, but such fortune cannot be expected generically and
a solution of that kind, if there is any, would be physically of little importance.
Hence we conclude here that there is no physically
relevant asymptotically dS black hole solution with the nontrivial
dilaton field under our assumptions in our model
except for the exact solutions which we will give in the following sections.
As a result, the black hole solutions exist only in the asymptotically AdS
spacetime.

\begin{figure}[t]
\begin{center}
\includegraphics[width=4cm]{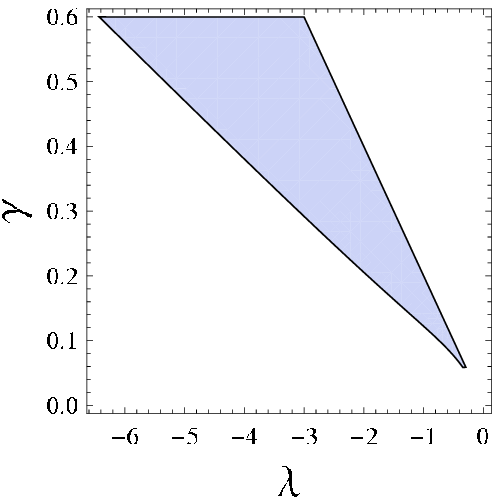}~~
\put(-60,-15){(a)}
\includegraphics[width=4cm]{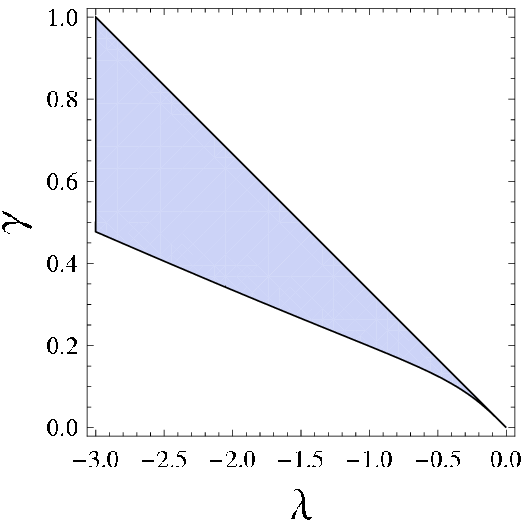}~~
\put(-60,-15){(b)}
\includegraphics[width=4cm]{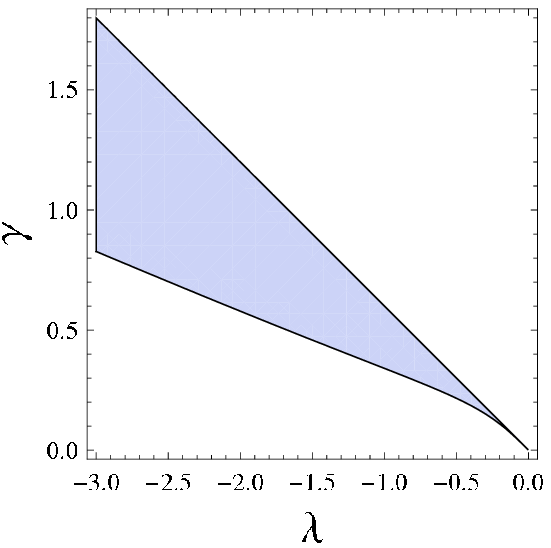}~~
\put(-60,-15){(c)}
\end{center}
\caption{
The allowed regions (shaded areas) for the parameters $\gamma$ and $\lambda$
for (a) $D=5$, (b) $D=6$, and (c) $D=10$.
The upper boundary comes from that the asymptotically AdS boundary condition,
and the lower boundary comes from
the BF bound condition~\p{condition_1}.
}
\label{allowed-parameter}
\end{figure}

For asymptotically AdS case $\gamma< (D-4)|\lambda|/D$, we impose the conditions as
in our previous paper,~\cite{GOT2}
\begin{equation}
\tm_{BF}^2\leq\tm^2 <0.
\label{condition_1}
\end{equation}
The allowed parameter region  $(\la,\c)$  of the conditions are depicted
in Figs.~\ref{allowed-parameter}. In $D=4$ there is no allowed parameter region.
This can be confirmed by Eq.~\p{class}.

\section{Hyperbolic topological black hole solutions with $k=-1$}

We now present our numerical solutions for $k=-1$ with a positive cosmological term.
We integrate the field equations from the event horizon to AdS infinity.
The first step in the procedure is to choose appropriate radius of the event
horizon~$\tilde{r}_H$.
Next we choose the value of $\phi_H$ and determine the values
of other fields by Eqs.~\p{bhor} and \p{pder}.
Since the asymptotic behavior \p{behavior} is in general not satisfied
for most of the values $\phi_H$, it should be tuned such that $\phi_-=0$ is achieved.
We also fix $\delta_H=0$ in the integration and this would give nonzero $\d_0$.
However, $\delta_0=0$ is always realized by the shift symmetry \p{sym3}.
As a result, there is only one freedom of choosing $\tilde{r}_H$,
given a cosmological constant.
The solutions are obtained for the particular choice of $\gamma$ and $\lambda$,
but we expect that qualitative properties do not change for other choices of
these parameters unless $\c$ is too large,~\cite{CGO1} though there is an indication
that the range of the horizon radii
for the existence of the black hole solutions changes depending on the strength
of the dilaton coupling $\c$~\cite{MOS}. In this section we fix $\gamma=1/2$.
Using the symmetry in Eq.~\p{sym2}, the solutions for different values of
cosmological constant can be generated from a solution for $\tilde\Lambda=+1$.
Indeed, solutions for $\tilde{\Lambda}_1$ can be obtained by simply changing
the variables as
\begin{equation}
\tilde{r} \to |\tilde{\Lambda}_1|^{-\frac{\gamma}{2(\lambda-\gamma)}}\tilde{r},
~~~~~
\phi \to \phi-\frac{1}{\lambda-\gamma}\log|\tilde{\Lambda}_1|,
~~~~~
\tilde{m}_g \to |\tilde{\Lambda}_1|^{-\frac{(D-3)\gamma}{2(\lambda-\gamma)}}\tilde{m}_g.
\end{equation}

In the following subsections, we present numerical solutions in various dimensions.
Before doing that, let us consider the analytical solution.
It can be confirmed that the basic equations have the exact solution
\bea
\phi \equiv \phi_0,~~
\delta \equiv 0,~~
B =  \tilde{b}_2 \tr^2-1,
\label{AdSBH}
\ena
where the relation of the parameters $b_2$, $\phi_0$ and the cosmological constant
are given as Eqs.~\p{b2_inf} and \p{phi_inf_2}.
This solution has an event horizon at $\tr=1/\sqrt{\tilde{b}_2}$, which coincides
with the AdS radius $\ell_{\rm AdS}$. By Eq.~(\ref{mass}), the mass of the black hole
solution is $\tm_g\equiv \tilde{M}_0 =0$. Hence this solution is the zero mass
black hole. The center of the solution is not singular but regular.

\begin{figure}[t]
\begin{center}
\includegraphics[width=4cm]{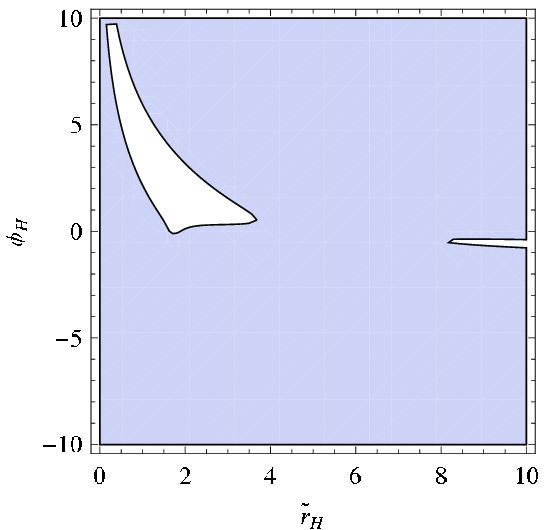}~~
\put(-60,-15){(a)}
\includegraphics[width=4cm]{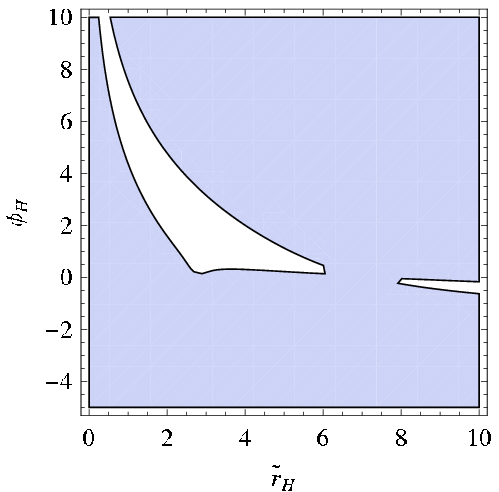}~~
\put(-60,-15){(b)}
\includegraphics[width=4cm]{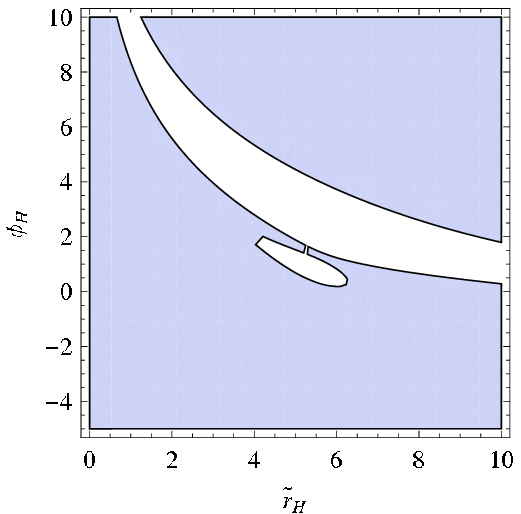}
\put(-56,-15){(c)}
\end{center}
\caption{
The region where Eq.~(\ref{pder}) has real roots of $\phi_H'$ (shaded area)
for $\tilde{\Lambda}=1$, $\gamma=1/2$, and $k=-1$ in (a) $D=5$ ($\lambda=-3$),
(b) $D=6$ ($\lambda=-2$),  and (c) $D=10$ ($\lambda=-1$).
On the boundary $\phi_H'$ degenerates. Only in this region, regularity at the event
horizon can be satisfied.
}
\label{f4}
\end{figure}

Let us examine the boundary condition \p{pder} in more detail here.
It is a quadratic equation with respect to $\phi_H'$. To guarantee the reality of
$\phi_H'$, we find that there are forbidden parameter regions of ($\phi_H'$, $\tr_H$).
Such regions (or allowed regions) are depicted in Fig.~\ref{f4} for our choice of
parameters $\gamma$, $\lambda$ and $\tilde\Lambda$ in each dimension.
The solutions can exist only in those shaded regions.
This does not, however, mean that there are always solutions for those values
in these regions. We also mention that the regions seem to change depending on
the parameter choice of $\c$ and $\la$,~\cite{MOS} but we will not discuss
this issue in this paper.

\begin{figure}[thb]
\begin{center}
\includegraphics[width=4cm]{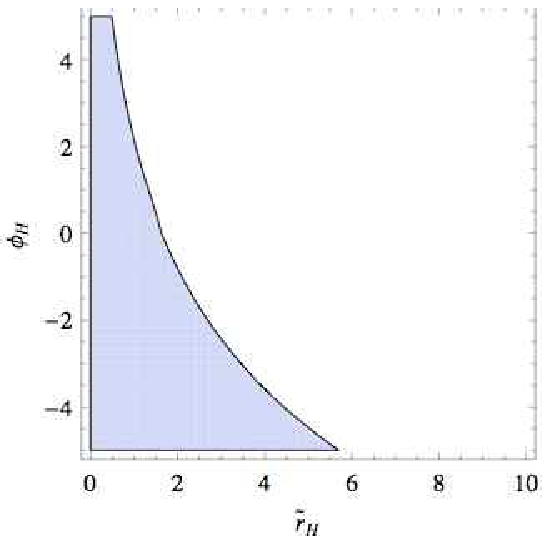}~~
\put(-60,-15){(a)}
\includegraphics[width=4cm]{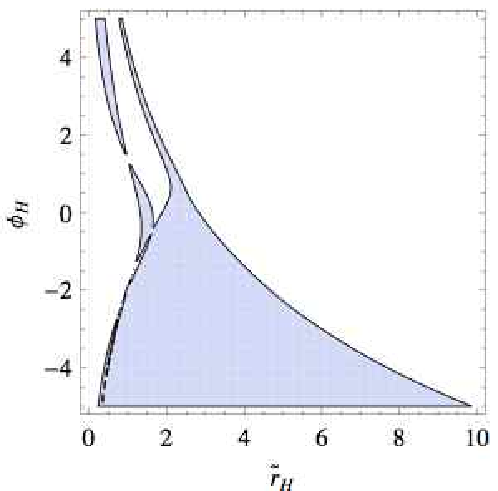}~~
\put(-60,-15){(b)}
\includegraphics[width=4cm]{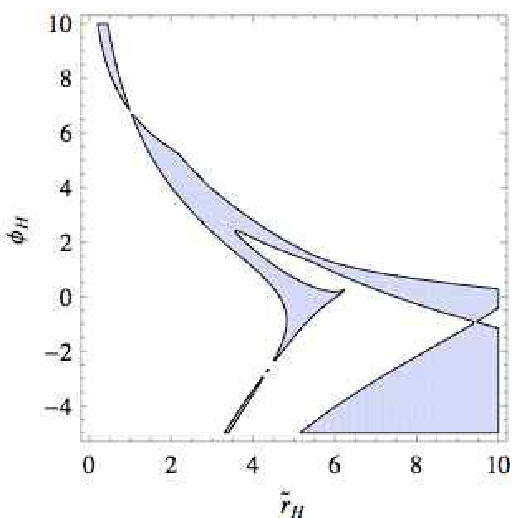}
\put(-60,-15){(c)}
\\
\includegraphics[width=4cm]{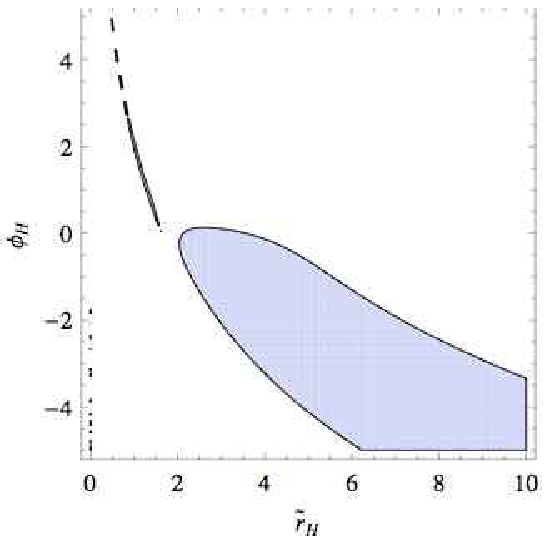}~~
\put(-60,-15){(a$'$)}
\includegraphics[width=4cm]{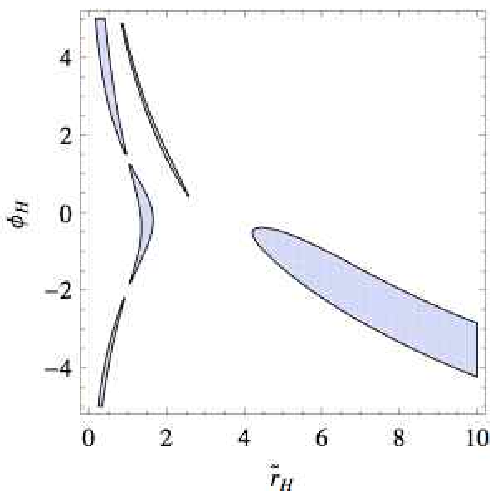}~~
\put(-60,-15){(b$'$)}
\includegraphics[width=4cm]{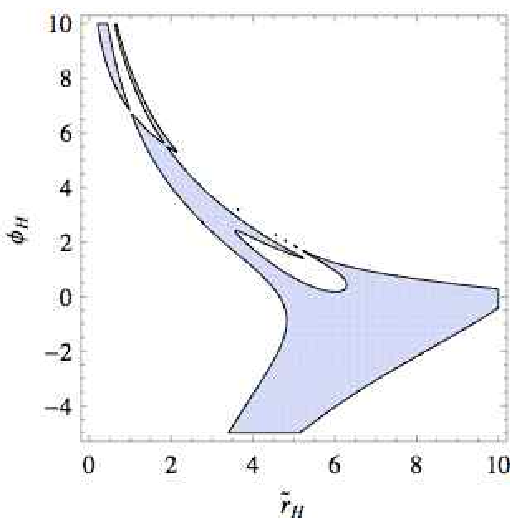}
\put(-60,-15){(c$'$)}
\end{center}
\caption{
The region where $B'_H>0$ for each solution of $\phi'_H$
for  $\tilde{\Lambda}=1$, $\gamma=1/2$, and $k=-1$ in (a), (a$'$) $D=5$ ($\lambda=-3$),
(b), (b$'$) $D=6$ ($\lambda=-2$), and (c), (c$'$) $D=10$ ($\lambda=-1$).
}
\label{f5}
\end{figure}

There is another boundary condition (\ref{hor}) which should be considered
when we choose the value of $\phi_H$.
The region where $B'_H>0$ is shown in Fig.~\ref{f5}. Outside of the shaded region,
the horizon, if it exists, is of cosmological horizon type.

\begin{figure}[htb]
\begin{center}
\includegraphics[width=15cm]{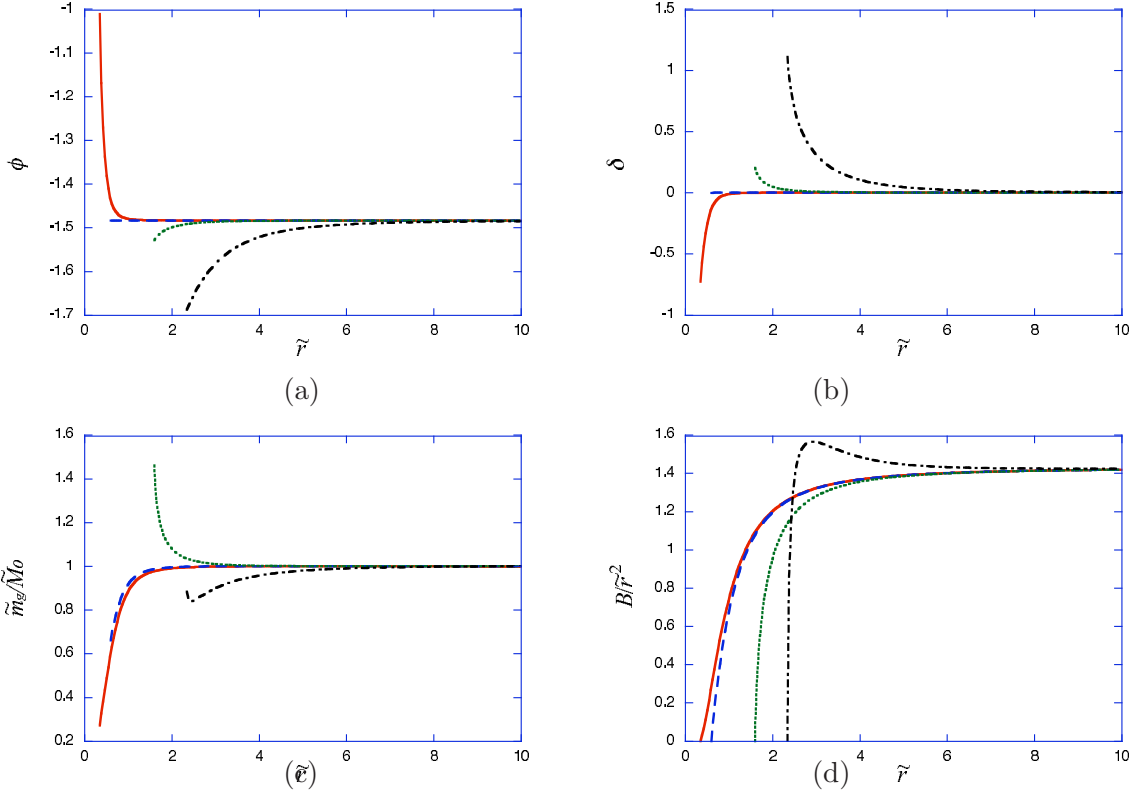}
\put(-320,145){\small (a)}
\put(-120,145){\small (b)}
\put(-320,0){\small (c)}
\put(-120,0){\small (d)}
\end{center}
\caption{
Behaviors of (a) the dilaton field $\phi(\tr)$, (b) the lapse function $\delta(\tr)$, 
(c) the mass function $\tm_g(\tr)/\tilde{M}_0$, and (d)
the metric function $B$ of the black hole
solutions with $k=-1$ in $D=5$. The horizon radii and the masses are
$\tr_H =0.35$, $\tilde{M}_0=-0.18614$ (solid line),
$\tr_H =0.6$, $\tilde{M}_0=-0.13238$ (dashed line),
$\tr_H=1.6$, $\tilde{M}_0=2.3287$ (dotted line),
and  $\tr_H=2.35$, $\tilde{M}_0=21.523$ (dot-dashed line).
}
\label{d5_config}
\end{figure}
\begin{figure}[htb]
\begin{center}
\includegraphics[width=15cm]{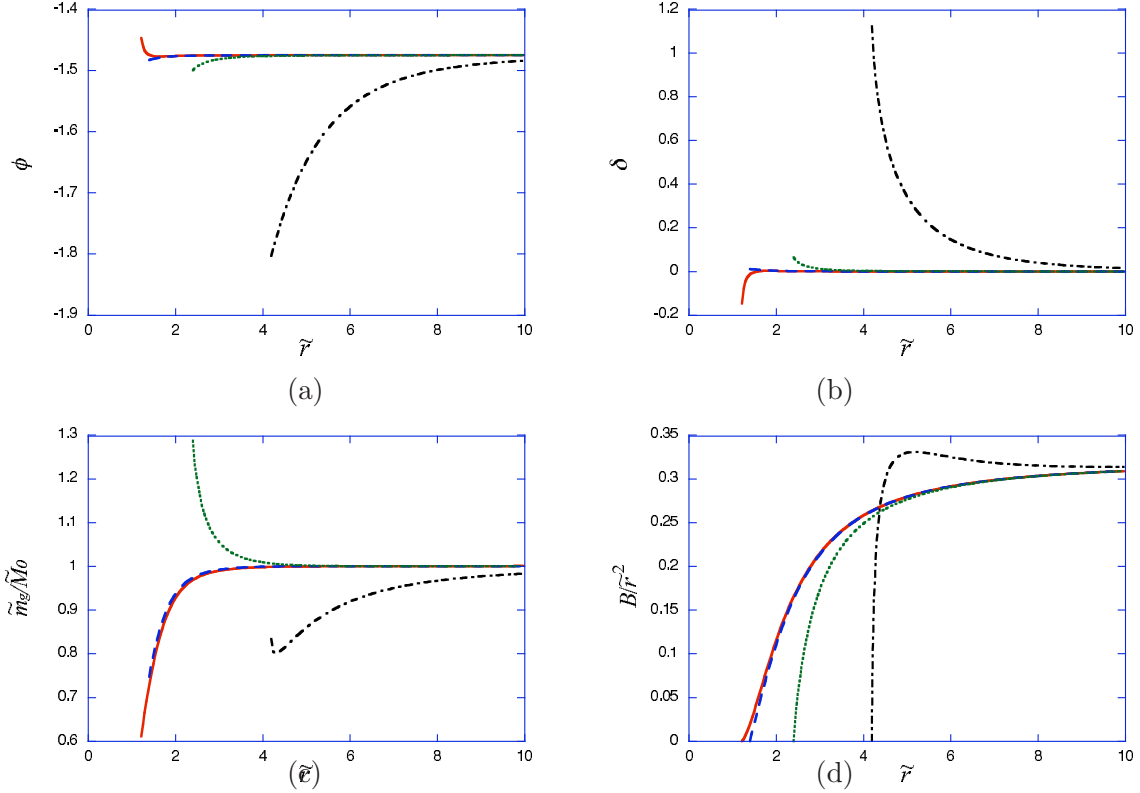}
\put(-320,145){\small (a)}
\put(-120,145){\small (b)}
\put(-320,0){\small (c)}
\put(-120,0){\small (d)}
\end{center}
\caption{
Behaviors of (a) the dilaton field $\phi(\tr)$, (b) the lapse function $\delta(\tr)$, 
(c) the mass function $\tm_g(\tr)/\tilde{M}_0$, and (d)
the metric function $B$ of the black hole
solutions with $k=-1$ in $D=6$. The horizon radii and the masses are
$\tr_H =1.22$, $\tilde{M}_0=-0.78098$ (solid line),
$\tr_H =1.4$, $\tilde{M}_0=-0.68890$ (dashed line),
$\tr_H=2.4$, $\tilde{M}_0=4.4924$ (dotted line),
and  $\tr_H=4.2$, $\tilde{M}_0=205.39$ (dot-dashed line).
}
\label{d6_config}
\end{figure}

\begin{figure}[htb]
\begin{center}
\includegraphics[width=15cm]{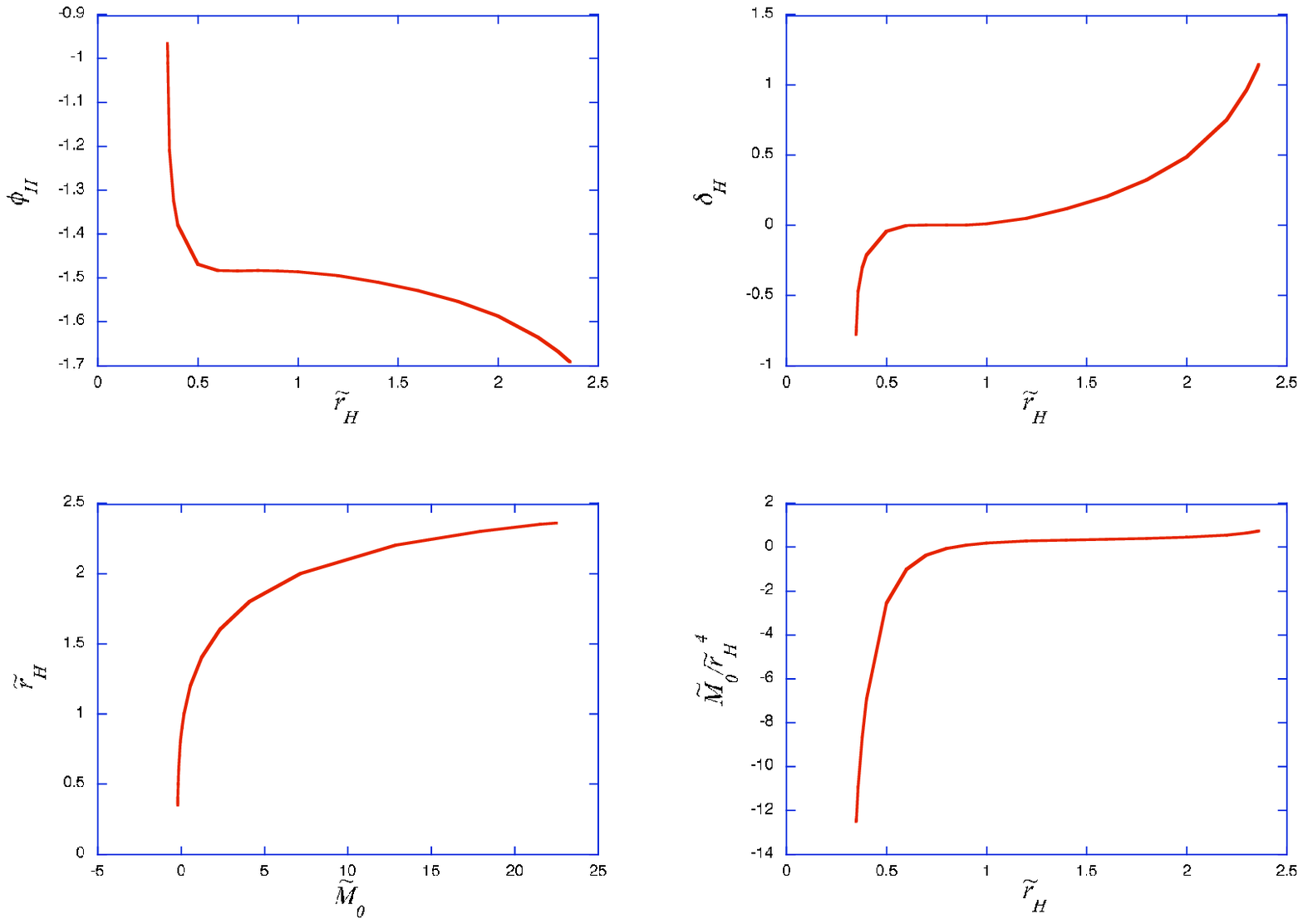}
\put(-320,145){\small (a)}
\put(-120,145){\small (b)}
\put(-320,0){\small (c)}
\put(-120,0){\small (d)}
\end{center}
\caption{
Numerical values of the fields (a) $\phi_H$ and (b) $\delta_H$ at the horizon
and (c) the mass of the black hole $\tilde{M}_0$-$\tr_H$ diagram for $D=5$ and $k=-1$.
(d) shows the scaled mass by horizon radius.
}
\label{d5_parameter}
\end{figure}

\begin{figure}[htb]
\begin{center}
\includegraphics[width=15cm]{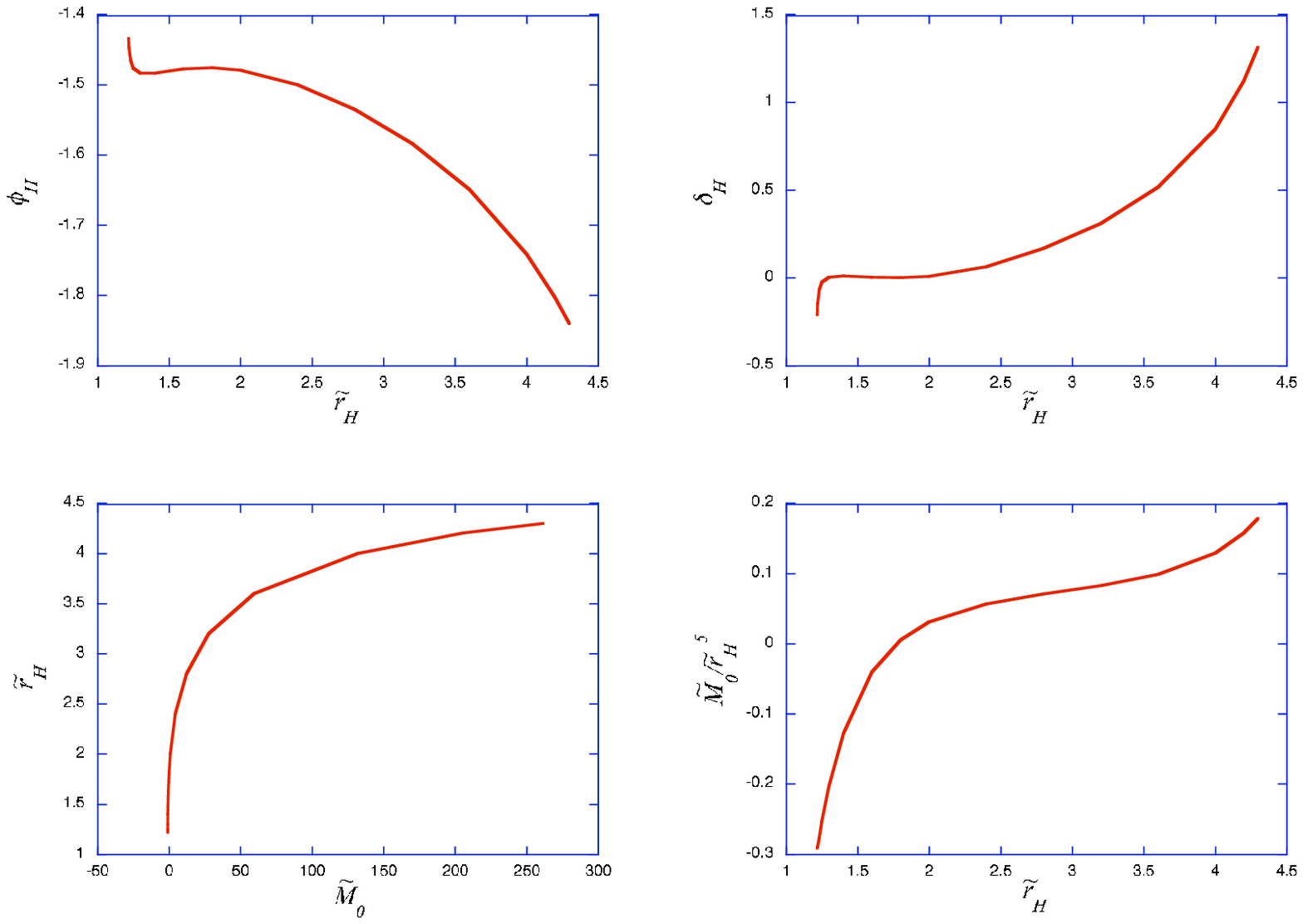}
\put(-320,145){\small (a)}
\put(-120,145){\small (b)}
\put(-320,0){\small (c)}
\put(-120,0){\small (d)}
\end{center}
\caption{
Numerical values of the fields (a) $\phi_H$ and (b) $\delta_H$ at the horizon
and (c) the mass of the black hole $\tilde{M}_0$-$\tr_H$ diagram for $D=6$ and $k=-1$.
(d) shows the scaled mass by horizon radius.
}
\label{d6_parameter}
\end{figure}

\subsection{$D=5$ and $D=6$ solutions}

First we present the black hole solutions in $D=5$ and $D=6$ simultaneously
since the qualitative features of these solutions are similar.
As the typical choice of parameters, we fix $\lambda=-3$ for $D=5$
and $\lambda=-2$ for $D=6$.
These values are chosen in the allowed region given in Fig.~\ref{allowed-parameter}.
We find from Eqs.~(\ref{b2_inf}) and (\ref{phi_inf_2}) that the square of inverse
AdS radius and the asymptotic value of the dilaton field are $\tilde{b}_2=1.42869$
($\ell_{\rm AdS}=0.8366$) and
$\phi_0=-1.4837$ for $D=5$, and $\tilde{b}_2=0.318784$ ($\ell_{\rm AdS}=1.7711$) and
$\phi_0=-1.47555$ for $D=6$, respectively.

The configurations of the field functions of the black hole solutions are
depicted in Fig.~\ref{d5_config} ($D=5$) and Fig.~\ref{d6_config} ($D=6$).
For the black holes with larger mass than that of the exact solution whose
horizon radius is $\tr_H=\ell_{\rm AdS}$, the dilaton field increases monotonically
to its asymptotic value $\phi_0$. Although the mass function decreases near
the event horizon, it increases asymptotically.
The function $\delta$ monotonically decreases.
As the radius of solution becomes large, we see that $B$ becomes steep around the
black hole horizon.
For the small black holes $\tr_H=0.6$ ($D=5$) and  $\tr_H=1.22$ ($D=6$),
the dilaton field decreases for the domain where $\tr \lesssim \ell_{\rm AdS}$.
The mass function decreases in most of the spacetime region although it
increases asymptotically (Note that $m_g(r)/M_0$ increases while $m_g(r)$
itself decreases when $M_0$ is negative). The function $\delta$ monotonically
increases.
Hence the qualitative features of the solutions change for $\tr_H \approx \ell_{\rm AdS}$.

The plots of $\phi_H$, $\delta_H$, and $\tilde{M}_0$ as functions of $\tr_H$
are given in Fig.~\ref{d5_parameter} ($D=5$) and Fig.~\ref{d6_parameter} ($D=6$).
It is noted that the black hole solution exists only for $0.3496<\tr_H<2.36$ ($D=5$)
and $1.218<\tr_H<4.3$ ($D=6$).

In the non-dilatonic case, there are also bounds for the existence of the black
hole solutions. In $D=5$, the lower bound is zero horizon radius where a naked
singularity appears, and the upper bound is determined by the appearance of
the branch singularity at finite radius\cite{TM}. In $D=6$, the lower bound is the extreme
black hole solution where the horizons degenerate, and the upper bound is
the branch singularity as in $D=5$. In the dilatonic case,
for the solution at the lower bound, the second derivative of the dilaton field $\phi''$
diverges at the horizon while $B'_H$ and $\delta_H$ are finite.
For the solution at the upper bound, $h_H$ becomes zero and $B'_H$ diverges.
The Hawking temperature is given by the periodicity of the Euclidean time on the
horizon as
\bea 
\label{temp}
\tilde{T}_H ~&& =\frac{e^{-\d_H}}{4\pi}B_H'
\nonumber \\
&& =\frac{e^{-\d_H}}{4\pi h_H}
\biggl[\frac{(D-3)k}{\tr_H} +\frac{(D-3)_5k^2}{\tr_H^2} e^{-\c\phi_H}
- \frac{ \tr_H\tilde\Lambda e^{\la\phi_H}}{D-2}\biggr].
\ena 
Hence the temperature of the black hole in the upper bound becomes infinite.
If we try to add the mass to the black hole to get the larger black hole than
upper bound, the emission rate of the black hole is extremely large and
the black hole may remain smaller than the upper bound.
The qualitative behavior of the physical quantities in Figs.~\ref{d5_parameter}
and \ref{d6_parameter} seems to be divided by zero mass exact solution (\ref{AdSBH}).
The gravitational mass $\tilde{M}_0$ is monotonic with respect $\tr_H$.
For the large black holes $\tilde{M}_0\sim \tr_H^{D-1}$ as in the non-dilatonic
case, while it grows faster for the small black holes.

\subsection{$D=10$ solution}

\begin{figure}[h]
\begin{center}
\includegraphics[width=15cm]{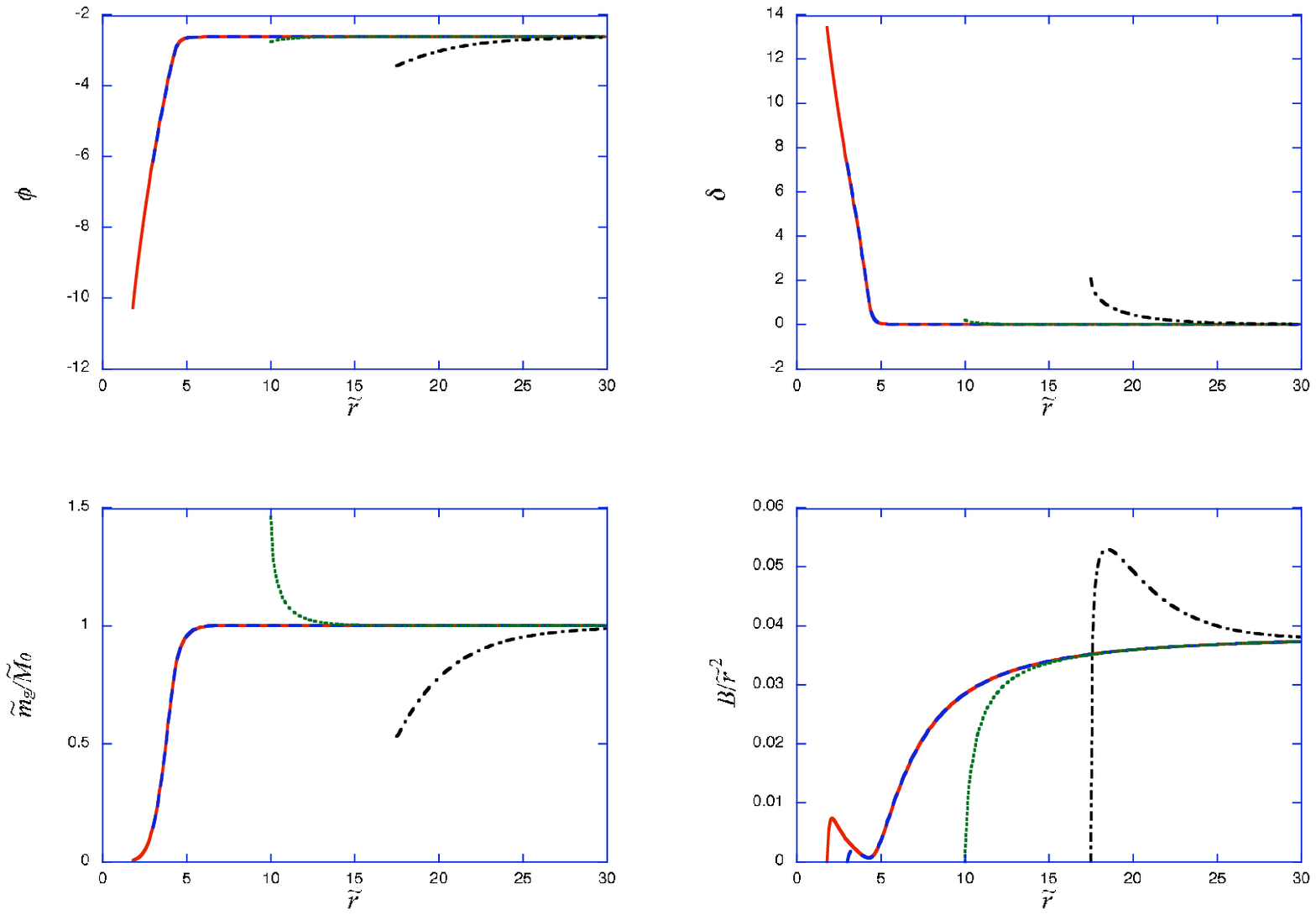}
\put(-320,145){\small (a)}
\put(-120,145){\small (b)}
\put(-320,0){\small (c)}
\put(-120,0){\small (d)}
\end{center}
\caption{
Behaviors of (a) the dilaton field $\phi(\tr)$, (b) the lapse function $\delta(\tr)$, 
(c) the mass function $\tm_g(\tr)/\tilde{M}_0$, and (d)
the metric function $B$ of the black hole
solutions with $k=-1$ in $D=10$. The horizon radii and the masses are
$\tr_H =1.8$, $\tilde{M}_0=-4919.2$ (solid line),
$\tr_H =3.0$, $\tilde{M}_0=-4909.1$ (dashed line),
$\tr_H=10.0$, $\tilde{M}_0=9.7349\times 10^{6}$ (dotted line),
and  $\tr_H=17.5$, $\tilde{M}_0=5.1025\times 10^{9}$ (dot-dashed line).
}
\label{d10_config}
\end{figure}

\begin{figure}[h]
\begin{center}
\includegraphics[width=15cm]{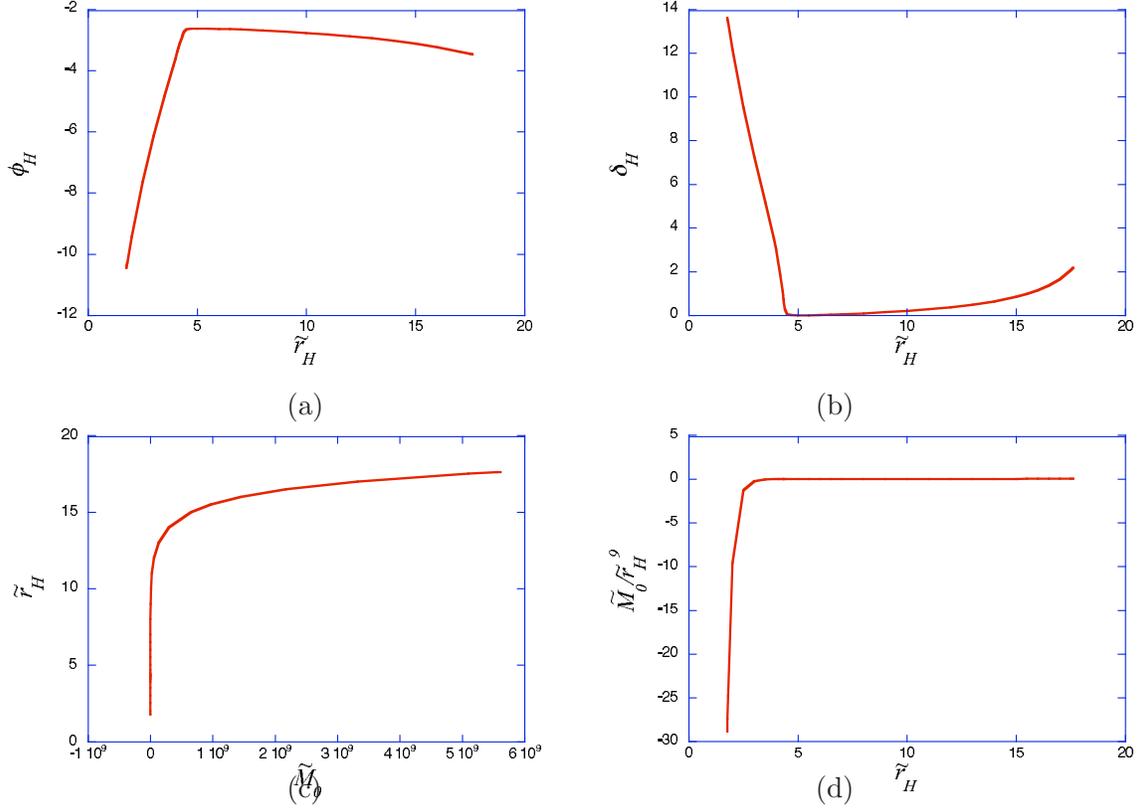}
\put(-320,145){\small (a)}
\put(-120,145){\small (b)}
\put(-320,0){\small (c)}
\put(-120,0){\small (d)}
\end{center}
\caption{
Numerical values of the fields (a) $\phi_H$ and (b) $\delta_H$ at the horizon
and the mass of the black hole (c) $\tilde{M}_0$-$\tr_H$ diagram in $D=10$.
(d) shows the scaled mass by horizon radius.
For each figure, $k=-1$.
}
\label{d10_parameter}
\end{figure}

\begin{figure}[h]
\begin{center}
\includegraphics[width=15cm]{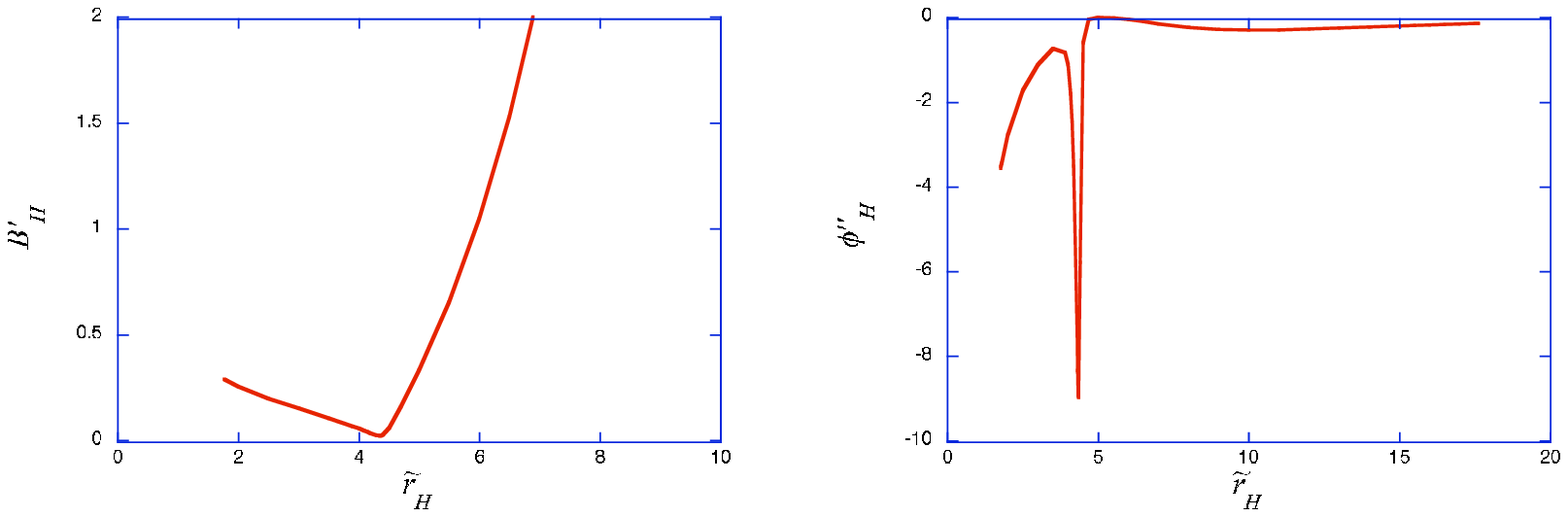}
\put(-320,0){\small (a)}
\put(-120,0){\small (b)}
\end{center}
\caption{
Numerical values of the fields (a) $B'_H$ and (b) $\phi''_H$ in $D=10$.
For each figure, $k=-1$.
}
\label{d10_para}
\end{figure}

The properties of the black hole solutions in 10-dimensional spacetime are
different from those in $D=5$ and $6$ cases. We choose $\lambda=-1$.
From Eqs.~(\ref{b2_inf}) and (\ref{phi_inf_2}), we find that
the square of inverse AdS radius and the asymptotic
value of the dilaton field are $\tilde{b}_2=0.038415$ ($\ell_{\rm AdS}=5.1021$) and
$\phi_0=-2.6268$, respectively.

The configurations of the field functions of the black hole solutions are given
in Fig.~\ref{d10_config}.
For the solutions with $\tr_H=10.0$ and $17.5$,
the dilaton field increases monotonically to its asymptotic value $\phi_0=0$.
The mass function for $\tr_H=17.5$ monotonically increases while it decreases
near the event horizon and increases asymptotically  for smaller black holes.
As the solution becomes large, we find that $B$ becomes steep around the event
horizon. The functions $\phi$, $\delta$, and $m_g$ have almost the same
configurations for the solutions with $\tr_H=1.8$ and $3.0$, i.e.,
the functions for $\tr_H=1.8$ and $3.0$ trace the same curves on the figures.
The metric function $B$ approaches zero at the local minimum around
$\tr=\tr^{\ast}\approx 4.36$ ($B'$ becomes zero there) and
bounces to $B/\tr^2 \to \tilde{b}_2$.
So the solution looks like extreme black hole from outside.
This is the common property for the solutions with horizon
radius smaller than $\tr^{\ast}$.

We show $\phi_H$, $\delta_H$, and $\tilde{M}_0$
as functions of $\tr_H$ in Fig.~\ref{d10_parameter}.
The black hole solution exists only for limited range of
horizon radius $1.76<\tr_H<17.61$.
The $\phi_H$-$\tr_H$ plot and  $\delta_H$-$\tr_H$ plot have cusp like structures
at the horizon radius $\tr_H=\tr^{\ast}$. This radius is smaller than the AdS
radius $\ell_{\rm AdS}=5.1021$.
When the horizon radius is larger than $\tr^{\ast}$, the solutions have similar
properties to those of the large black holes in $D=5$ and $6$.
$\phi_H$ monotonically decreases and $\tilde{M}_0\sim \tr_H^{D-1}$ asymptotically.
For the largest $\tr_H=17.61$, $B'_H$ diverges, which means that the
temperature of the black hole also diverges.
For solutions with $\tr_H$ smaller than  $\tr^{\ast}$, the solutions have different
properties from large one. $\phi_H$ and $\delta_H$ seem to be linear to $\tr_H$.
The solution disappears at the lower bound $\tr_H=1.76$, where the non-normalizable
mode of the dilaton field cannot be eliminated by tuning the shooting parameter $\phi_H$.

For the horizon radius $\tr_H=\tr^{\ast}$, the value $B'_H$ vanishes
(See Fig.~\ref{d10_para} (a)), which means that the horizon degenerates.
However, this is not an extreme black hole solution with a regular horizon
because the second derivative of the dilaton field
diverges (See Fig.~\ref{d10_para} (b)).

\section{Planar topological black hole solution with $k=0$}

The basic equations have the following exact solution
\bea
\phi \equiv \phi_0,~~
\delta \equiv 0,~~
B =  \tilde{b}_2 \tr^2,
\ena
where the relation of the parameters $b_2$ and $\phi_0$
are again given as Eqs.~\p{b2_inf} and \p{phi_inf_2}.
For $\gamma>(D-4)|\lambda|/D$, $\tilde{b}_2$ is negative and
the metric function $B$ is negative semidefinite. So this solution is
physically less interesting.
For $\gamma<(D-4)|\lambda|/D$, $\tilde{b}_2$ is positive.
The spacetime is regular everywhere and approaches the AdS spacetime
asymptotically.

It can be proved simply that there is no black hole solution with $k=0$.
In the $k=0$ case, Eq.~\p{bhor} becomes
\bea
B_{H}'  = -  \frac{\tr_H\tilde\Lambda e^{\la\phi_H}}{D-2}<0.
\ena
This means that the horizon is not the black hole horizon but the cosmological
horizon type which is not the solution we are looking for.
Hence there is no black hole solution in this case.

\section{Spherically symmetric black hole solutions with $k=1$}

It can be confirmed that the basic equations have the exact solution
\bea
\phi \equiv \phi_0,~~
\delta \equiv 0,~~
B =  \tilde{b}_2 \tr^2 + 1,
\label{dS}
\ena
where the relation of the parameters $b_2$ and $\phi_0$
are again given as Eqs.~\p{b2_inf} and \p{phi_inf_2}.
For $\gamma>(D-4)|\lambda|/D$, the solution is regular dS solution.
In the $k=-1$ case, there is also an exact solution (\ref{AdSBH}) which is
a zero-mass black hole solution. By adding static perturbations around that solution,
the black hole solutions with the nontrivial dilaton field and infinitesimal mass
are obtained. Similarly, it might be expected that the regular solution with
a nontrivial dilaton field is obtained by static perturbations in the $k=1$ case.
It could be a new nontrivial solution with a cosmological horizon.
At the cosmological horizon, the dilaton field should satisfy the regularity
condition, and the degree of freedom of the dilaton field at the center
is used to realize this.
Unfortunately then, the growing mode in the dilaton field that arises asymptotically
as discussed in \S~\ref{AllowedRegions} cannot be eliminated since the freedom
is already used up. There is certianly a possibility that this might happen to
vanish, but this is not expected in general. Hence such an asymptocic dS black
hole solution, if it could exist for very particular set of parameters,
may have little importance physically.

For $\gamma<(D-4)|\lambda|/D$, the solution is AdS solution. Depending on
the parameters, the solution satisfies the BF bound. By adding static
perturbations to this exact solution, we may obtain an everywhere regular
solution with non-trivial dilaton field in the asymptotically AdS spacetime.
In this case, there is no cosmological horizon and we have the freedom to
kill the growing mode and obtain the good asymptotic AdS behavior.
Though this is not a black hole solution, it is an interesting solution in
its own right. We will report on this possibility in the near future.

As in the topological black hole case ($k=-1$), we made various choices of the
parameters of the theory $\gamma$ and $\lambda$, and integrated the field
equations from $\tr=\tr_H$ to see if there is an asymptotically AdS black hole solution.
The metric function $B$ grows from zero and takes local maximum
and then decreases. As $B$ approaches zero, the field equations become
singular, and $\phi$ and $B$ seem to diverge at finite radius. Thus, even if we succeed
in getting $B=0$ corresponding to the cosmological horizon by tuning $\phi_H$,
the spacetime does not approach AdS spacetime. This means that the spacetime is singular,
and  suggests that there is no such solution.
Having searched for improved solutions by changing the parameters of theory
$\gamma$, $\lambda$, the physical parameter $r_H$ and shooting parameter $\phi_H$,
we have not found any reasonable solution after long struggle.
Together with the above discussions on the general behaviors of the function $B$,
we conclude that it is very unlikely that there is any asymptotically AdS black
hole solution in $k=1$ case with positive cosmological constant.

\section{Hyperbolic topological solutions with $k=-1$ without cosmological constant}
\label{zeroLambda}

In this series of our papers, there still remains a case that has not been
investigated.
It is the case of $k=-1$ and zero cosmological constant. Here we briefly report
on our study of this case.

The basic equations are obtained by putting $k=-1$ and $\tilde{\Lambda}=0$
in Eqs.~(\ref{fe1})-(\ref{dil}). These equations have the same symmetry
properties and scaling rules as Eqs~(\ref{sym1})-(\ref{sym3}).
In particular, the shift symmetry (\ref{sym2}) is written as
\bea
\phi \to \phi-\phi_{\ast}, ~~
r \to e^{-\gamma\phi_{\ast}/2}r, ~~
\delta \to \delta, ~~
B \to B.
\label{sym4}
\ena
This can be used to shift the asymptotic value of the dilaton field to zero
even when we compute the metric functions and the dilaton field for any boundary
value $\phi_H$ at the horizon.
In other words, we can first fix the horizon radius, for example, $\tr_H=1$,
and obtain various solutions with different $\phi_H$. Then applying the shift symmetry
(\ref{sym4}), we can obtain the solutions with various horizon radii and zero
dilaton field at infinity.

In the non-dilatonic case, there are two branches of the solutions with $\Lambda=0$
and $k=-1$\cite{TM}. The GR branch does not give a black hole solution while
there are black hole solutions for some mass range in the GB branch. Such black hole
solutions have asymptotic AdS structure. If there is a black hole solution
in the dilatonic system, it cannot have AdS structure for the following reason.
In the AdS spacetime, the GB term becomes dominant asymptotically and it can be
regarded as the Liouville type potential of the dilaton field. Hence the potential
does not have extremum, and the dilaton field diverges. This means that a black hole
solution, if it exists, does not approach the non-dilatonic solutions
in the non-dilatonic limit ($\gamma\to 0$) and it should be new type of solutions.

\begin{figure}[tbh]
\begin{center}
\includegraphics[width=6cm]{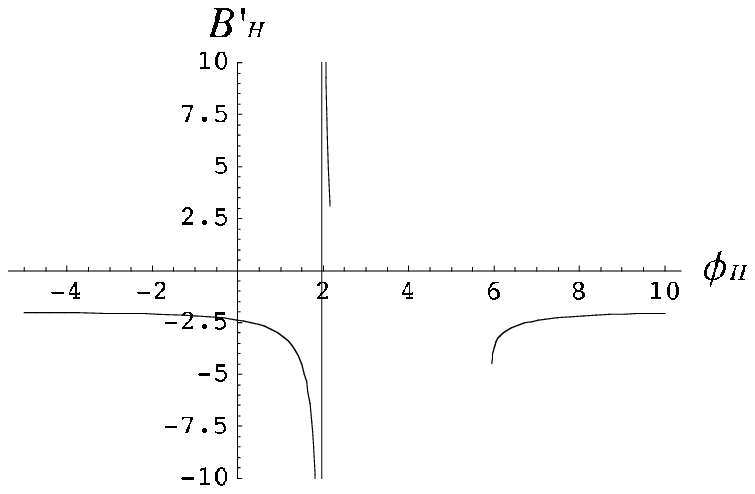}~~
\put(-90,-15){(a)}
\hspace{3mm}
\includegraphics[width=6cm]{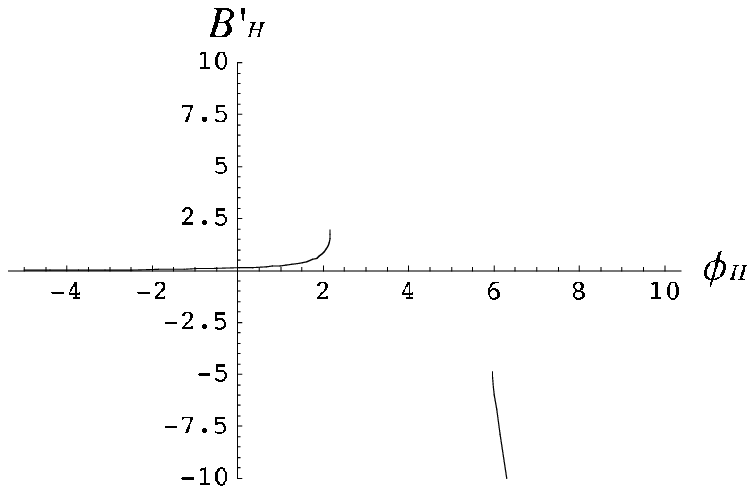}~~
\put(-90,-15){(a$'$)}
\\
\vspace{3mm}
\includegraphics[width=6cm]{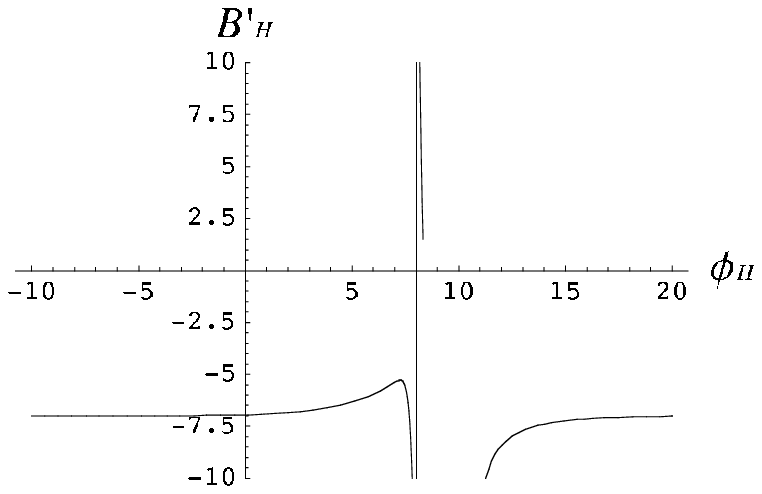}
\put(-90,-15){(b)}
\hspace{3mm}
\includegraphics[width=6cm]{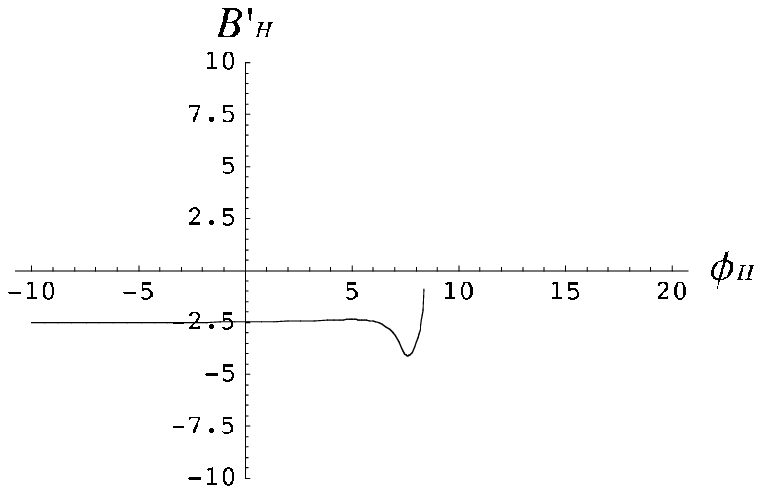}~~
\put(-90,-15){(b$'$)}
\end{center}
\caption{
The $B'_H$-$\phi_H$ plots for each solution
for  $\Lambda=0$, $\gamma=1/2$, $k=-1$, and $\tr_H=1$ in $D=5$ ((a) and (a$^{\prime}$))
and $D=10$ ((b) and (b$^{\prime}$)). Two plots in each dimension correspond to two roots for
$\phi'_H$ in Eq.~(\ref{pder}). The range for $\phi'_H$ with no plot implies
that $\phi'_H$ becomes imaginary.
}
\label{L0hyperbolic}
\end{figure}

We assume the same boundary conditions 1-3 in \S~\ref{BC}. At the horizon,
Eq.~(\ref{bhor}) becomes
\bea
B_{H}'  =-\frac{1}{h_H}\biggl[ \frac{(D-3)}{\tr_H} +\frac{(D-4)_5C}{2}\biggr].
\label{B_H}
\ena
This must be positive for the existence of the black hole horizon.
In $D=4$, Eq.~(\ref{B_H}) becomes
\bea
B_{H}'  =-\frac{1}{h_H \tr_H}.
\label{B_Hd4}
\ena
For $D=4$, Eq.~\p{pder} gives
\bea
h_H^2 - h_H + 6\c^2 C^2=0,
\ena
whose two solutions are, if they are real, both positive.
Hence $B'_H$ is always negative, excluding the existence of the black hole solution
in $D=4$.
In $D\geq 5$, Eq.~\p{pder} is too complicated to deduce definite conclusion.
So we fix the value of $\gamma$ and examine the positivity of $B'_H$.
Figs.~\ref{L0hyperbolic} are the plots of  $B'_H$ as a function of  $\phi_H$
for $\gamma=1/2$. $\tr_H$ is fixed to $\tr_H=1$ by the scaling
property of the system. We find there is a narrow range where $B'_H$ becomes positive.
However, integrating the field equations from $\tr=1$ with
$\phi_H$ in this range, we find that calculation stops, and the field variables
diverge. This gives a rather strong evidence that there is no black hole solution
in this case either.\footnote{Some solutions are found in the string frame
if we relax the condition that the dilaton is constant asymptotically.~\cite{MOS}
Because they satisfy different boundary conditions from ours, they do not
correspond to black holes we are looking for.
}

\section{Discussions and Conclusions}
\label{CD}

We have studied the black hole solutions in the dilatonic Einstein-GB theory with
the positive cosmological constant. The cosmological constant introduces
the Liouville type of potential for the dilaton field with a certain coupling.
We have studied the spherically symmetric ($k=1$),
planar symmetric ($k=0$), and hyperbolically symmetric
($k=-1$) spacetimes in various dimensions.

The black hole solution should have a regular black hole horizon
and be singularity free in the outer region.
Due to the presence of the positive cosmological constant, it is expected that the
cosmological horizon appears. To obtain the regular cosmological horizon,
the boundary value of the dilaton field at the black hole horizon should be
tuned. However, it is found by the asymptotic expansion at infinity
that the dilaton field has a non-normalizable mode for such solutions.
As a result, there is no physically relevant black hole solutions in asymptotically
dS spacetime generically.

In the planar symmetric spacetime, all the horizons become the cosmological
horizon type, i.e., $B'_H<0$, which means that there is no black hole solution.

In the spherically symmetric and the hyperbolically symmetric spacetimes,
there are parameter regions where the solutions can have asymptotically
AdS spacetimes.
By the asymptotic expansion at infinity, the power decaying rate of the fields
are estimated. We have imposed the condition that the ``mass"
of the dilaton field satisfies the BF bound, which guarantees the stability of
the vacuum solution. By this condition, the values of the dilaton coupling
constant and the parameter of the Liouville potential are constrained.

In the spherically symmetric spacetime, however, we could not find any solution
with AdS asymptotics numerically, and concluded that there is no black hole solution
with $k=1$ in this system.

On the other hand, we were able
to construct AdS black hole solutions in hyperbolically symmetric spacetime
in higher dimensions ($D\geq 5$) numerically.
The basic equations
have some symmetries which are used to generate the black hole solutions
with different horizon radii and the cosmological constants.
We have chosen $\gamma=1/2$ and
$\lambda=-3$ ($D=5$), $\lambda=-2$ ($D=6$), and $\lambda=-1$ ($D=10$),
for the actual numerical analysis.
The field equations have exact solutions, i.e., a massless black hole solution
with the constant dilaton field.

The black hole solutions in the $D=5$ and $D=6$ spacetimes have similar
properties. Configurations of the field functions change qualitatively
depending on whether the size of the black hole is larger than the AdS radius or not.
For the small black holes, the GB term becomes efficient, and the solutions
deviate from the non-dilatonic one.
The black hole solutions exist for a certain range of the size of horizon radius.
At the lower bound, the second derivative of the dilaton field diverges at the horizon,
and there appears a singularity at finite radius. This is similar to the
spherically symmetric black hole solution without the cosmological
constant.~\cite{GOT1,TYM}
At the upper bound, the derivative of the metric function $B'$ diverges at the
horizon. The temperature of the black hole becomes infinite.

In the $D=10$ spacetimes there is a special solution, which is singular, for the
horizon radius $\tr^{\ast}$. Since the derivative of the metric function $B'$ vanishes
at the black hole horizon, the solution looks similar to the extreme black hole
solution. However, it is not extreme one because the second derivative of the dilaton field
diverges.
When the horizon radius is larger than $\tr^{\ast}$, the solutions have similar
properties to the large solutions in lower dimensions $D=5$ and $6$. There is
an upper bound for the horizon radius where $B'_H$ diverges.
When the horizon radius is smaller than  $\tr^{\ast}$, it is interesting that
the configurations of the field functions trace the similar curves to those of
the solution with the horizon radius $\tr_H=\tr^{\ast}$ outside of $\tr^{\ast}$.
The solutions disappear at the lower bound, where the non-normalizable mode
of the dilaton field cannot be eliminated by tuning the value of $\phi_H$.

The characteristic feature of the solutions in this model is the existence of the
upper bound for the size of the black holes. If the matter which has more mass
than the upper bound, i.e., macroscopic size compared to the string scale,
collapses gravitationally, the final state is not the static
black hole but something else. One of the candidates is the appearance of the naked
singularity, which is an undesirable situation physically.
Another possibility is that the matter field does not collapse to form the black hole
and the spacetime remains regular everywhere like a soliton or a lump solution.
This is allowed since the GB term breaks the energy condition if it is
regarded as the ``matter part" in GR and the singularity theorem does not apply.
This is a kind of singularity avoidance.

We have also studied the hyperbolically symmetric black hole solutions without
the cosmological constant which was not investigated in our previous papers.
In the $D=4$ case, it is proved that all the horizon is of the cosmological
horizon type and that there is no black hole solution.
In the higher dimensional cases, it appears that the solution can have
a black hole horizon.
However, the parameter region for the suitable boundary conditions is
very narrow, and we did not find any black hole solution in this case.

Since this is the final paper of our series about the dilatonic black hole solutions
in various dimensions, it is helpful to summarize the
results obtained in our papers.
We show some properties of the solutions in Table~\ref{table_sammary}
in  $D=4,\:5, \:6,$ and 10 spacetimes.
In the $\Lambda=0$ case, there is a black hole solutions only for spherically
symmetric ($k=1$) spacetime. The solutions are asymptotically flat and there is
the lower bound for the size of the horizon in $D=4,5$, and $6$ but there is not such
a bound in $D=10$.\footnote{In our papers, we choose the value of the
coupling constant $\gamma$ corresponding to the value in $D=10$. It was pointed out
that the existence of the lower bound for the horizon radius depends on
its value.~\cite{MOS} There is this ambiguity in $\c$ because it changes depending
on whether we first go to the Einstein frame in $D=10$ and then make dimensional
reduction to lower dimensions or first make dimensional reduction and then go to
the Einstein frame.
}

\begin{table}
\begin{center}
\caption{
Summary of the black hole solutions in the dilatonic Einstein-Gauss-Bonnet theory
in various dimensions.
The head ``existence" shows the existence of the black hole solution which satisfy
appropriate boundary conditions. For the $\Lambda=0$, $k=-1$ and the $\Lambda=1$, $k=1$
cases, ``no" means that the non-existence of the solution cannot be proved
exactly but we cannot find the solution numerically.  For the $\Lambda=1$, $k=-1$
case, there are solutions for $D=5,6$, and 10 but there is no solution for $D=4$.
The  head ``asymptotics" shows the asymptotic structure of the solution.
There is no asymptotically dS solution generically.
The head ``lower bound" and ``upper bound" show the existence of the lower and
upper bounds for the horizon radius.   For the $\Lambda=1$, $k=1$ case,
there is the lower bound for $D=4,\:5$ but no bound for $D=6,\:10$.
The head ``paper" shows the number in the series of our paper where
the model is discussed.
}
\label{table_sammary}
\begin{tabular}{c|c||ccccc}
\hline
     & $k$     & existence & asymptotics & lower bound & upper bound & paper \\
\hline
                       &   &  &    &  yes ($D=4,\: 5$)  &    &   \\
                       & \raisebox{1.5ex}[0pt]{$k=1$}
                       & \raisebox{1.5ex}[0pt]{yes} 
                       & \raisebox{1.5ex}[0pt]{flat}   
                       & no ($D=6,\:10$)  
                       & \raisebox{1.5ex}[0pt]{no}   
                       & \raisebox{1.5ex}[0pt]{I}  \\
\cline{2-7}
\raisebox{1.5ex}[0pt]{$\Lambda=0$} & $k=0$  & no  & -----  & -----     & ----- & II  \\
\cline{2-7}
                       & $k=-1$ & ``no" & ----- & -----   & ----- & IV  \\
\hline
                       & $k=1$  & ``no" & ----- & -----   & -----& IV  \\
\cline{2-7}
  & $k=0$  & no   & ----- & -----   & ----- & IV  \\
\cline{2-7}
 \raisebox{1.5ex}[0pt]{$\Lambda=1$} & &  yes ($D\geq 5$) & & & & \\
 &  \raisebox{1.5ex}[0pt]{$k=-1$} 
 &  no ($D=4$)
 &  \raisebox{1.5ex}[0pt]{AdS}
 &  \raisebox{1.5ex}[0pt]{yes} 
 & \raisebox{1.5ex}[0pt]{yes}
 & \raisebox{1.5ex}[0pt]{IV}  \\
\hline
             & $k=1$  & yes & AdS & yes      & no   & III  \\
\cline{2-7}
$\Lambda=-1$ & $k=0$ & yes & AdS & no       & no   & II  \\
\cline{2-7}
             & $k=-1$ & yes & AdS & yes      & no   & III  \\
\hline
\end{tabular}
\end{center}
\end{table}

For the positive cosmological term, the black hole solutions exist only for
hyperbolically symmetric ($k=-1$) and $D=5,\:6,\:10$ spacetimes and
there is no black hole solution in $D=4$.
We have obtained the allowed parameter region of $\gamma$ and $\lambda$ for
the existence of the solutions.
In spite of the positive cosmological constant,
the spacetime is asymptotically AdS. There are both the upper and lower
bounds on the size of the black hole.

For the negative cosmological term, there are black hole solutions for all $k$.
We have found the allowed parameter region of $\gamma$ and $\lambda$ for
the existence of the solutions.
They are asymptotically AdS and do not have upper bound for the size of
the event horizon. There is no lower bound on the horizon for the $k=0$ solutions
while there is for the $k=\pm1$ solutions as in the $\Lambda=0$ and $k=1$ case.

Non-existence of the black hole solution in asymptotically dS spacetime
is strongly related to the cosmological horizon. The equation of the dilaton field
becomes singular at horizons, and regularity of the horizons gives
constraints on the dilaton field and its derivative. To satisfy these constraints,
the dilaton field has to have suitable value at each horizons. There is another
boundary condition, i.e., the asymptotic behavior at infinity. This gives a further
constraint on the dilaton field. However, there is no freedom to satisfy all the
constraints at more than one horizon and at infinity simultaneously.
This is the basic reason why there is no asymptotically dS black hole
solution in this system. In general, it is expected that the system of the matter field
whose equation of motion is singular at the horizons cannot have a
black hole solution in asymptotically dS spacetime for a similar reason.

For the non-dilatonic black holes in the Einstein-Gauss-Bonnet theory, global structures
of all the static solutions with $k=\pm 1$ and $k=0$ are investigated\cite{TM}.
Since our numerical analysis was limited to outer spacetime of the event horizon,
the global structures of the solutions such as the existence of the inner horizon,
and the singularity have not been clarified.
This is left for future study.

The stability of our solutions is another important subject.~\cite{PC}
For the $\Lambda=0$ and $k=1$ case in $D=4$, the stability against
the spherically symmetric perturbations is examined\cite{torii}.
In the non-dilatonic case, it was found that the black hole solutions are
stable against the spherical perturbations but can be unstable against
the non-spherical perturbations for the small solutions\cite{DG,Soda}.
It is interesting to see whether the effects of the dilaton field can stabilize the
solutions.


\section*{Acknowledgements}
We would like to thank Kei-ichi Maeda for valuable discussions.
N.O. was supported in part by the Grant-in-Aid for
Scientific Research Fund of the JSPS Nos. 20540283,
and also by the Japan-U.K. Research Cooperative Program.
T.T. was supported in part by the Grant-in-Aid for
Scientific Research Fund of the MEXT No. 21740195.

\appendix


\end{document}